\documentclass[twocolumn,twocolappendix]{aastex63}

\newcommand\pp{\hat P}
\newcommand\rrho{\hat\rho}
\newcommand\uu{\hat u}

\received{\today}
\revised{...}
\accepted{...}
\submitjournal{The Astrophysical Journal}

\usepackage{mathrsfs}  
\usepackage{needspace}
\usepackage{amsmath}

\def\note #1]{{\bf #1]}}
\def\dd{{\rm d}}

\newif\ifref
\reftrue
\reffalse
\definecolor{darkerred}{rgb}{0.5, 0, 0}
\definecolor{red}{rgb}{1, 0, 0}
\newcommand{\mb}[1]{\ifref\textcolor{darkerred}{#1}\else #1\fi}
\newcommand{\mbb}[1]{\ifref\textcolor{darkerred}{#1}\else #1\fi}

\newif\ifreff
\refftrue
\refffalse
\newcommand{\mbbb}[1]{\ifreff\textcolor{darkerred}{#1}\else #1\fi}

\newif\ifrefff
\reffffalse
\newcommand{\mbbbb}[1]{\ifrefff\textcolor{darkerred}{#1}\else #1\fi}

\shorttitle{Asteroseismic Inference of the Central Structure in a Subgiant Star}
\shortauthors{Bellinger et al.}
\graphicspath{{./}{figures/}}

\begin{document}

\title{Asteroseismic inference of the central structure in a subgiant star}

\correspondingauthor{Earl P.\ Bellinger}
\email{bellinger@phys.au.dk}

\author[0000-0003-4456-4863]{Earl P.\ Bellinger}
\affiliation{Stellar Astrophysics Centre, 
Department of Physics and Astronomy, 
Aarhus University, 
Denmark}

\author[0000-0002-6163-3472]{Sarbani Basu}
\affiliation{Department of Astronomy, 
Yale University, 
New Haven, CT, USA}

\author[0000-0002-1463-726X]{Saskia Hekker}
\affiliation{Center for Astronomy of Heidelberg University (Landessternwarte), Heidelberg, Germany}
\affiliation{Heidelberg Institute for Theoretical Studies, Heidelberg, Germany}
\affiliation{Max Planck Institution for Solar System Research, 
G\"ottingen, Germany}
\affiliation{Stellar Astrophysics Centre, 
Department of Physics and Astronomy, 
Aarhus University, 
Denmark}

\author[0000-0001-5137-0966]{J{\o}rgen Christensen-Dalsgaard}
\affiliation{Stellar Astrophysics Centre, 
Department of Physics and Astronomy, 
Aarhus University, 
Denmark}

\author[0000-0002-4773-1017]{Warrick H.\ Ball}
\affiliation{School of Physics and Astronomy,
University of Birmingham, 
UK}


\begin{abstract}
Asteroseismic measurements enable inferences of the underlying stellar structure, such as the \mb{density and the} speed of sound at various points within the interior of the star. 
This provides an opportunity to test stellar evolution theory by assessing whether the predicted structure of a star agrees with the measured structure. 
Thus far, this kind of inverse analysis has only been applied to the Sun and three solar-like main-sequence stars. 
Here we extend the technique to stars on the subgiant branch, and apply it to one of the best-characterized subgiants of the \emph{Kepler} mission, HR~7322. 
The observation of mixed oscillation modes in this star facilitates inferences of the conditions of its inert helium core, nuclear-burning hydrogen shell, and the deeper parts of its radiative envelope. 
We find that despite significant differences in the mode frequencies, the structure near to the center of this star does not differ significantly from the predicted structure. 
\end{abstract} 


\section{Introduction} \label{sec:intro}
The Sun and Sun-like stars pulsate with standing acoustic waves, the precise measurement of which has yielded a treasure of insight into their global and interior properties \citep[e.g.,][]{2010aste.book.....A, 2016LRSP...13....2B, basuchaplin2017, 2019LRSP...16....4G, 2019arXiv191212300A}. 
By comparing measurements of stellar oscillation frequencies to those from stellar evolution models, it is possible to tightly constrain their ages and evolutionary properties \mb{\citep[e.g.,][among numerous other examples]{1984srps.conf...11C, 1994ApJ...427.1013B, 2014A&A...569A..21L, 2015MNRAS.452.2127S, 2019A&A...622A.130B, 2019MNRAS.484..771R, 2020MNRAS.499.2445H}}. 
These investigations assume the truth of evolutionary theory, and attribute the ages of the best-fitting models to the star. 
However, models from stellar evolution simulations do not match asteroseismic or even helioseismic data perfectly. 
As stellar oscillation frequencies depend on the underlying stellar structure, this indicates that the internal structures of the models are not exactly right. 

Asteroseismology provides the opportunity to assess the quality of stellar models by offering a window into the internal structure of a star, and thus the means to test stellar evolution theory by comparing the star's structure with the structure of the best-fitting stellar evolution models. 
This kind of analysis, called a structure inversion, has revealed most of the solar interior \citep[e.g.,][]{1985Natur.315..378C, 1996Sci...272.1296G, 1996ApJ...460.1064B, 2009ApJ...699.1403B, 2016LRSP...13....2B, 2020A&A...642A..36B, 2020arXiv200706488C} as well as the core structure of three solar-like main-sequence stars \citep[][]{2017ApJ...851...80B, 2019ApJ...885..143B}. 
These efforts have led to major proposed revisions to the standard solar model \citep[e.g.,][]{2019ApJ...881..103Z} as well as the suggestion that mixing processes may be missing from evolutionary models of main-sequence stars \citep{2019ApJ...885..143B}. 

\begin{figure}
    \centering
    \includegraphics[width=\linewidth]{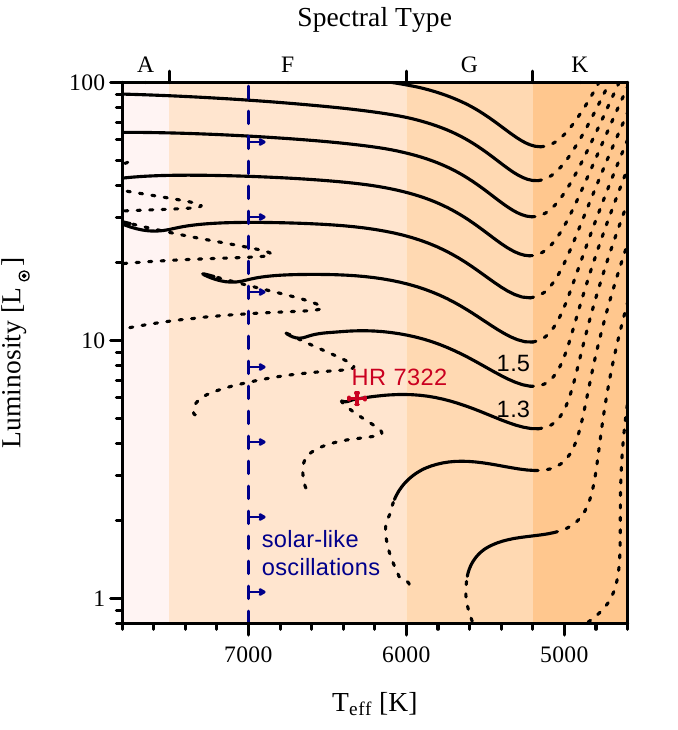}
    \caption{Hertzsprung--Russell diagram showing the predicted evolution of stars with masses of $0.7 - 2.9~M_\odot$ with diffusion and overshooting. The masses of two tracks are indicated. The subgiant phase is drawn with solid lines; the main sequence and red giant branch phases are in dotted lines. Stars cooler than approximately 7000~K, such as HR~7322, have convective envelopes and consequently are observed to exhibit solar-like oscillations. Models obtained from MIST \citep{2016ApJS..222....8D, 2016ApJ...823..102C}. \label{fig:hr}  } 
\end{figure}

Following the cessation of hydrogen fusion in their central regions, yet prior to becoming red giants, solar-like stars enter a relatively brief stage of stellar evolution known as the subgiant branch (see Figure~\ref{fig:hr}). 
In this phase, hydrogen fusion in a shell outside of the core supplies the star with energy as its core contracts and its outer layers expand and cool. 
As a consequence of its evolving structure, the acoustic oscillations that are excited in the outer regions of the star successively take on the behavior of internal $g$-mode oscillations that are otherwise invisibly trapped within the interior \mb{\citep{1975PASJ...27..237O, 1977A&A....58...41A}}. 
These mixed oscillation modes provide insight into the deep internal structure of the star, and present the opportunity to measure the sound speed at a depth even surpassing that of the Sun. 

\mbb{In this work, we present an investigation of the effectiveness of an inverse analysis of oscillation data for subgiants. 
We first describe the background and method in detail, and present blind tests using models. 
We then apply these methods to the subgiant HR~7322 \citep{2019MNRAS.489..928S}. 
We conduct the inversions using four different reference models in order to study the systematic errors of the procedure.} 

\mbb{Recently, \citet{2020IAUS..354..107K} have performed an inverse analysis on two subgiant stars. 
They claimed to resolve the density profile throughout the core, shell, and inner parts of the envelope of these models. 
They also found that they were able to resolve the isothermal sound speed profile only outside of the core.
In contrast, we find that we are able to infer the sound speed of the inert helium core as well as in the hydrogen burning shell and in the deeper parts of the radiative envelope for HR~7322. 
Furthermore, we are able to infer the density of the core, though not elsewhere in the star.} 

\begin{figure}
    \centering
    \includegraphics[width=\linewidth,trim={0 0 0 0cm}, clip]{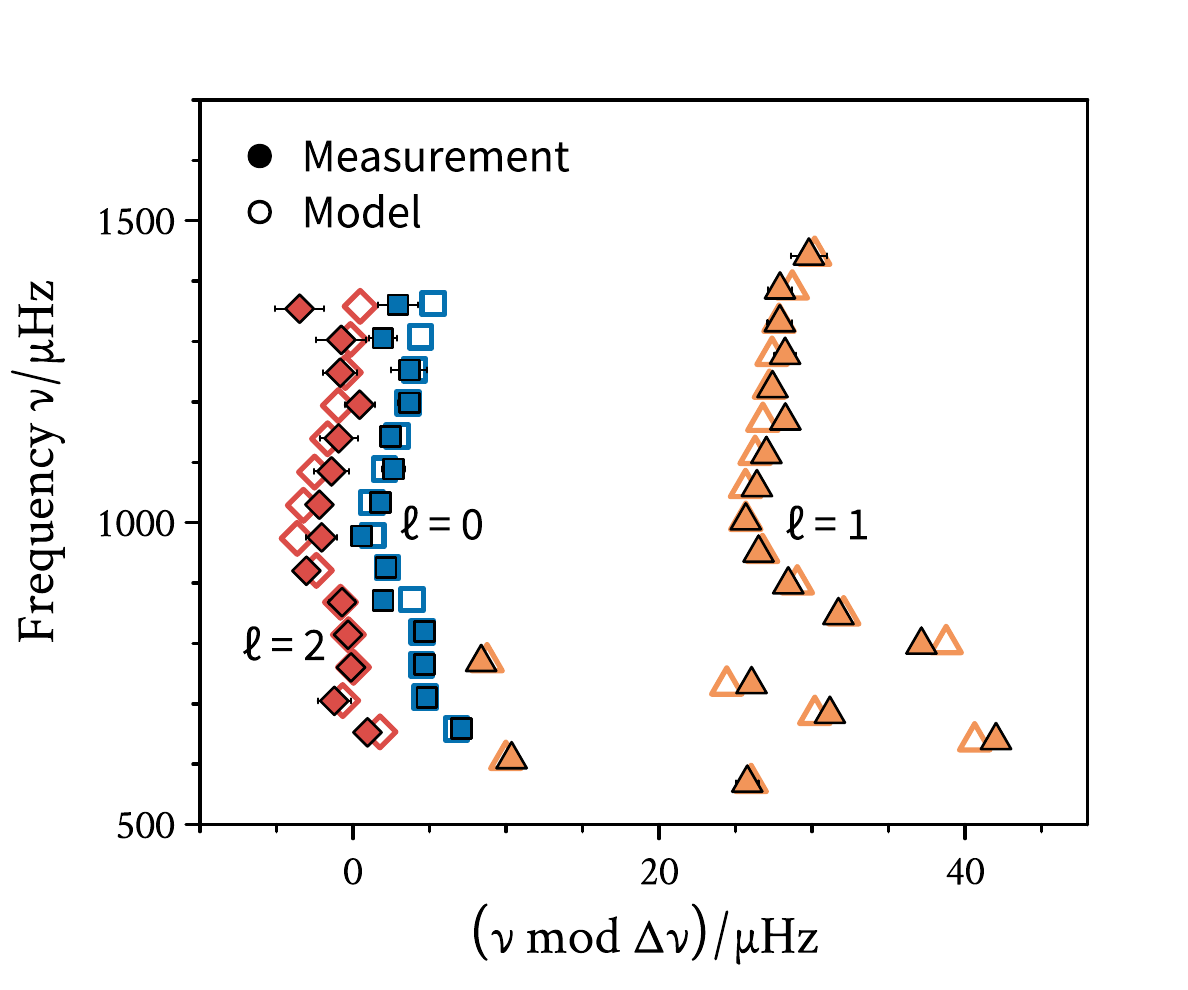}%
    \caption{
        \'{E}chelle diagram of HR~7322 and a best-fitting evolutionay model whose frequencies have been corrected for surface effects using the \citet{2014A&A...568A.123B} two-term correction. 
        A low-frequency avoided crossing is apparent in the dipolar modes with $\nu\sim700\,\mu\text{Hz}$. 
        Frequency measurement uncertainties at one standard deviation are shown, although in most cases they are smaller than the symbol size. 
        Though the fit of the model is qualitatively good, in the sense that absolute differences are small, significant differences are apparent, meaning that the structure of the model is not the same as that of the star.  \label{fig:echelle}} 
\end{figure}

The target of our analysis, HR~7322, was observed in short-cadence mode by the \emph{Kepler} spacecraft \citep{2010Sci...327..977B} during its nominal mission, as well as by the optical interferometer CHARA \citep{2005ApJ...628..453T} and the high-resolution spectrograph SONG \citep{2014RMxAC..45...83A}, making it one of the best-characterized subgiants to date. 
An \'{e}chelle diagram comparing the mode frequencies of HR~7322 to an evolutionary model is shown in Figure~\ref{fig:echelle}. 
It is apparent that even after correcting for the inadequate modeling of the near-surface layers, significant differences remain between the model and the star.
These differences are the starting point of this study. 

\needspace{2\baselineskip}
\section{Stellar Structure and Oscillations}
The frequencies of stellar oscillations are determined by the structure of the star. 
Perturbations to the stellar structure therefore result in perturbations to the stellar oscillation frequencies. 
This fact can be used to infer the differences in internal structure between a star and a suitably chosen model, known as the reference model, by analyzing the differences in their frequencies.

From a linear perturbation analysis of the equations of non-adiabatic stellar oscillations \citep[e.g.,][]{1991sia..book..519G, 1993ASPC...40..541G, 1999JCoAM.109....1K, basuchaplin2017}, one can find equations that relate changes in the internal stellar structure to changes in the frequency of each oscillation mode: 
\begin{equation} \label{eq:inversion}
    \frac{\delta \sigma_i}{\sigma_i}
    =
    \int
    K_i^{(\hat u,Y)}\, \frac{\delta \hat u}{\hat u}
    +
    K_i^{(Y,\hat u)}\, \delta Y
    \;\text{d}x. 
\end{equation}
Here the index $i$ refers to the radial order $n$ and spherical degree $\ell$ of a given oscillation mode. 
The dimensionless frequency of the mode is denoted $\sigma_i$, and is related to the cyclic frequency $\nu$ by 
\begin{equation}
    \sigma^2 = \left( \frac{GM}{R^3} \right)^{-1} \left(2\pi\nu \right)^2
\label{eq:sigma}
\end{equation}
where $G$ is the gravitational constant, $R$ is the photospheric radius, and $M$ is the total stellar mass. 
Differences are denoted with the $\delta$ symbol, e.g., $\delta \sigma_i$ refers to the difference between the frequency of the $i$th mode before and after perturbation, or the difference between an observed mode frequency and the corresponding frequency of the reference model. 
Integration in Equation~(\ref{eq:inversion}) takes place over the fractional radius $x\equiv r/R$ of the star, where $r$ is the distance from the stellar center, and differences such as $\delta \hat u$ in structure variables are taken at fixed $x$. 

\begin{figure}
    \centering
    \includegraphics[width=\linewidth]{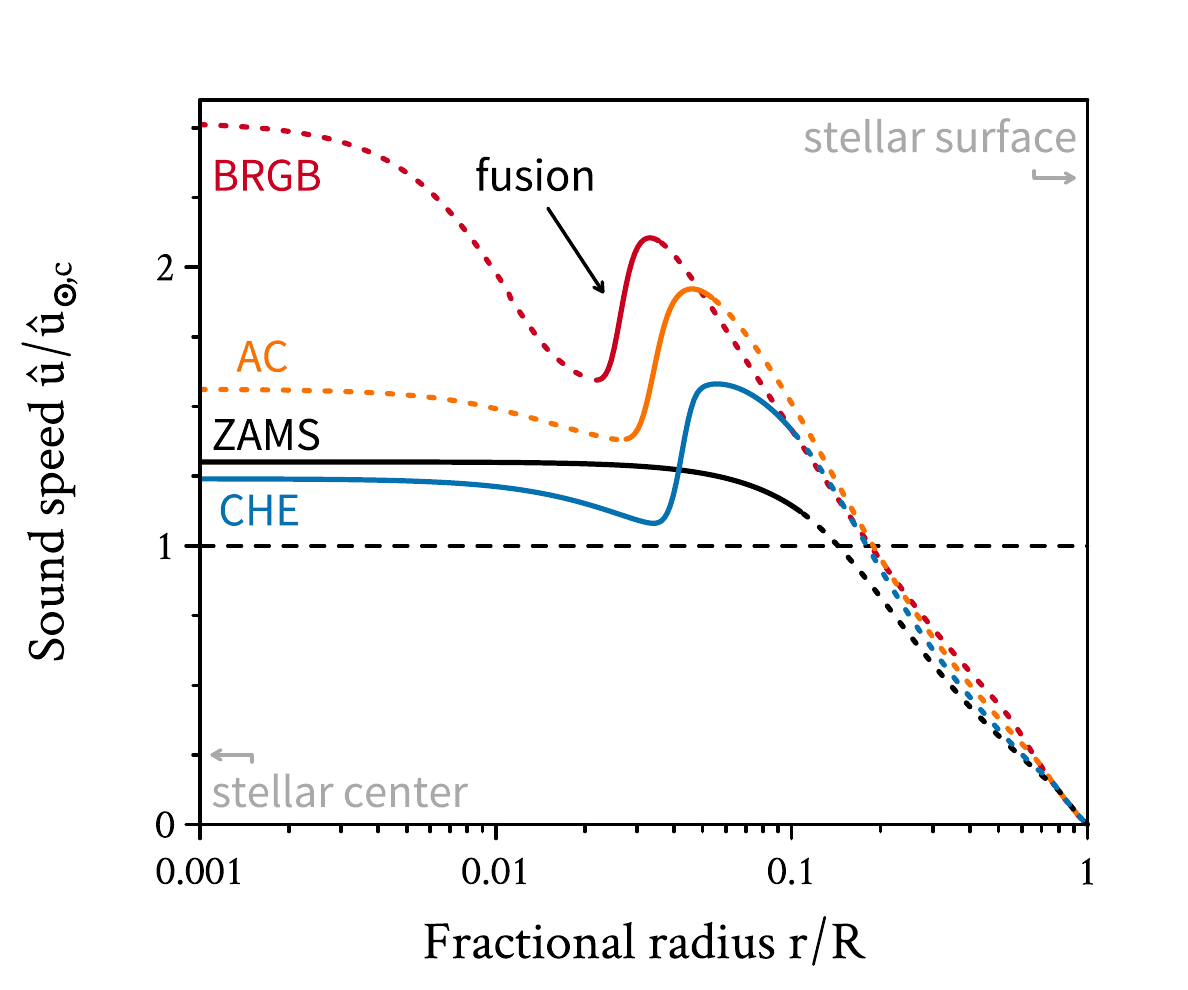}
    \caption{Predicted evolution of the acoustic structure (\emph{cf.}~Equation~\ref{eq:u}) of a $1.2~\text{M}_\odot$ star. 
    Lines show the structure of models at the zero-age main sequence (ZAMS), near core-hydrogen exhaustion (CHE), the first avoided crossing in the observable frequency spectrum (AC), and at the base of the red giant branch (BRGB). 
    Solid lines indicate regions of hydrogen fusion. 
    Values are normalized by the central dimensionless squared isothermal sound speed of the Sun, denoted $\hat u_{\odot,c}$. 
    \label{fig:soundspeed} } 
\end{figure}

The quantity $\hat u$ is the dimensionless squared isothermal sound speed, and is the principal factor in determining stellar oscillation frequencies. 
It varies throughout the stellar interior based on the local pressure $P$ and density $\rho$: 
\begin{equation} \label{eq:u}
    \hat u = \left( \frac{GM}{R} \right)^{-1} \frac{P}{\rho}.
\end{equation}
As the square of the sound speed is approximately proportional to the ratio of the temperature to the mean molecular weight of the stellar plasma, $\hat u$ varies over time as a consequence of the nuclear evolution of the star (see Figure~\ref{fig:soundspeed}). 
Asteroseismic measurements of $\hat u$ at one or more points within the interior of the star---and subsequent comparison with its theoretically predicted value at that location---is thus the goal of the analysis.

The fractional helium abundance at any location within the star is denoted $Y$, which increases internally as a result of hydrogen fusion, and decreases in the outer layers as a result of diffusion into the star. 
Other physics, perhaps including effects that are hitherto unknown, may also alter its abundance. 
However, despite being formally part of Equation~(\ref{eq:inversion}), $\delta Y$ is known to have little bearing on the oscillation frequencies outside of ionization zones \citep[e.g.,][]{2003Ap&SS.284..153B}. 
Nevertheless, we include it in our analysis. 
We do, however, neglect differences in metallicity, which play a smaller role than changes in $Y$ \citep{2017A&A...598A..21B}. 

The kernels $K$ are sensitivity functions derived from the perturbation analysis that link the differences in structure to the differences in mode frequencies, and are numerically computed from the chosen reference model \citep[e.g.,][]{2002ESASP.485...95T}. 
Some aspects of the kernels $K^{(\uu, Y)}$ and $K^{(Y, \uu)}$ are discussed in Appendix~\ref{sec:kerprop}.
\mb{Pairs of variables other than $(\uu,Y)$ may also be useful for asteroseismic inverison, such as $(\hat \rho, Y)$ \citep{2020IAUS..354..107K}, where $\hat\rho = R^3 M^{-1} \rho$. We explore this pair later as well.}

To illustrate some aspects of Equation~(\ref{eq:inversion}), we have simulated the evolution of a $1.2~\text{M}_\odot$ star and computed the adiabatic oscillation frequencies of each model throughout the subgiant phase of evolution. 
We have used both the \textsc{Mesa} stellar evolution code \citep[r12115\footnote{Run with the otherwise default settings of \texttt{use\_eosDT2} and \texttt{use\_eosELM} turned off, as these were found to produce erroneous equation of state values \citep{mesausers}.},][]{2011ApJS..192....3P, 2013ApJS..208....4P, 2015ApJS..220...15P, 2018ApJS..234...34P, 2019ApJS..243...10P} as well as the Aarhus Stellar Evolution Code \citep[\textsc{Astec},][]{2008Ap&SS.316...13C}. 
The conclusions were the same between the two codes, and hence for brevity we will here only show the \textsc{Mesa} models. 
We computed the adiabatic stellar oscillation frequencies using \textsc{Adipls} \citep[][]{2008Ap&SS.316..113C} and the kernels using an adapted form of the \texttt{kerexact} routines that were kindly provided to us by the late Michael J.\ Thompson (private communication).

\begin{figure}
    \centering
    \includegraphics[width=\linewidth,trim={0 0 0 0cm}, clip]{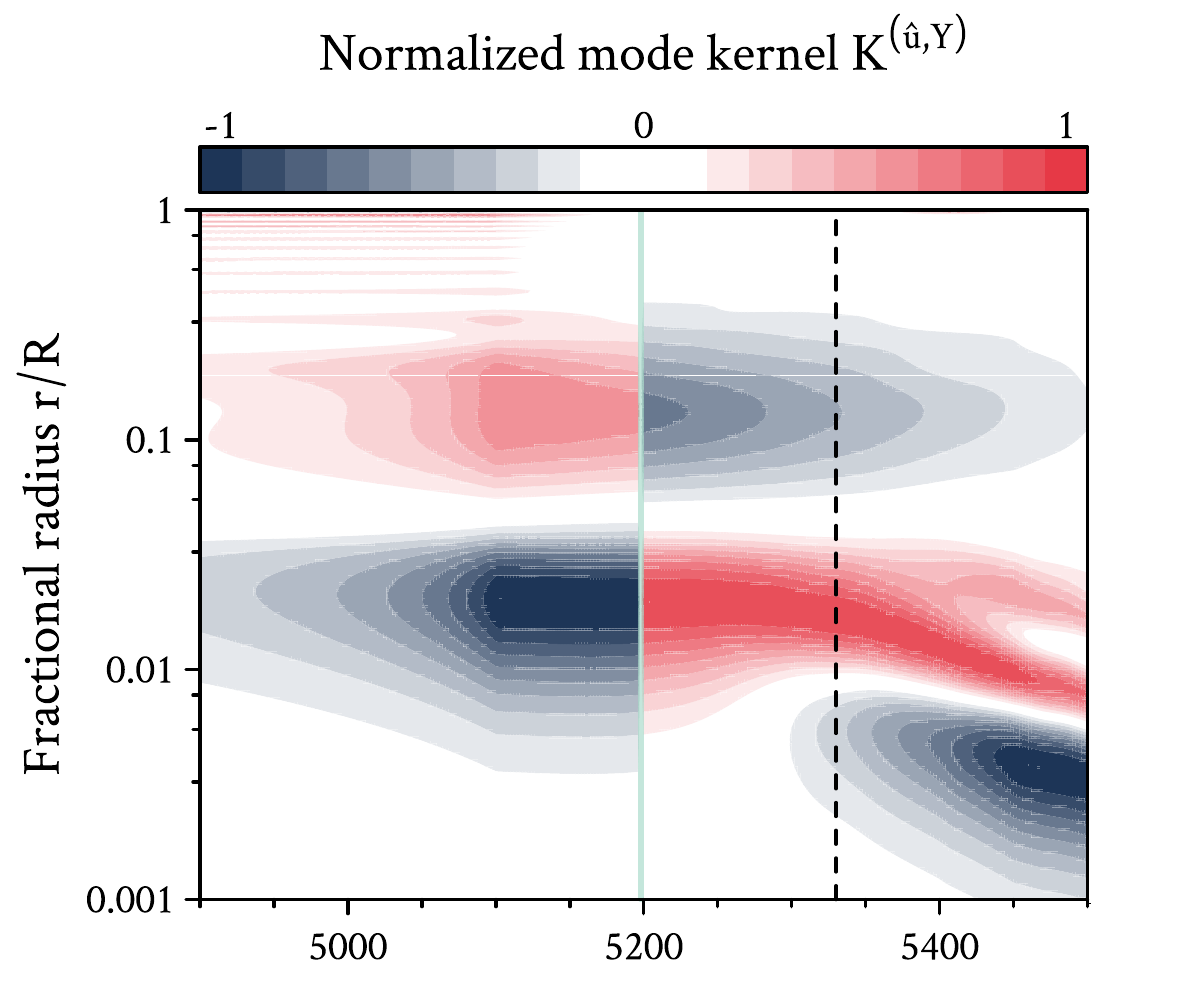}\\%
    \includegraphics[width=\linewidth,trim={0 0 0 1cm}, clip]{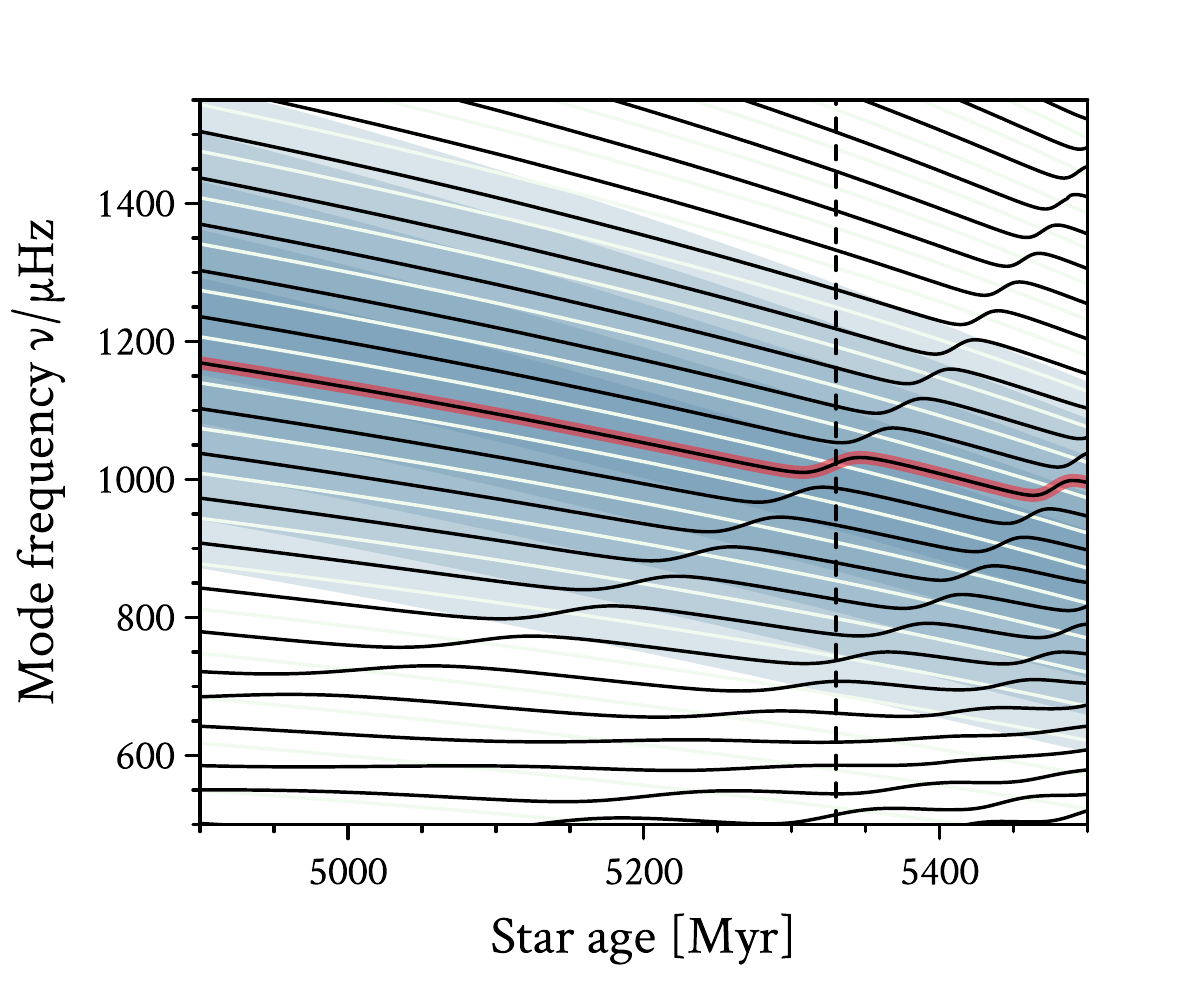}%
    \caption{
        \textsc{Top Panel}: Evolution of the $(\hat u,Y)$ kernel for the ${n=16}$ dipolar oscillation mode during the subgiant branch for a $1.2~\text{M}_\odot$ stellar track. 
        The kernels have been normalized by their maximum absolute value.
        A sign change is apparent at $\sim 5200$~Myr as indicated by the vertical green line (see also Appendix~\ref{sec:kerprop}).  
        The vertical dashed line indicates the first avoided crossing for that mode, after which point it becomes sensitive to the deep stellar interior. 
        \textsc{Bottom Panel}: Evolution of radial (white lines) and dipolar (black lines) oscillation mode frequencies for the same track. 
        The background shading indicates the distance from $\nu_{\max}$ in units of the large frequency separation $\Delta\nu$ (up to a maximum of $5$), which roughly indicates the region that is currently detectable in a well-observed star. 
        The otherwise uniform pattern of stellar oscillation frequencies as seen in the left side of the figure becomes visibly perturbed by the emergence of mixed oscillation modes. 
        The mode whose kernel is shown in the top panel is highlighted. 
        \label{fig:kernel-evol} 
        } 
\end{figure}

The evolution of the oscillation modes, in terms of both their frequencies as well as their structure kernels, are shown in Figure~\ref{fig:kernel-evol}. 
The sensitivity of the depicted dipolar mode to the central structure of the star increases enormously when it gains the character of a $g$-mode at an age of around 5350~Myr. 
A change of sign is also apparent on the subgiant branch around 5200~Myr. 
This corresponds to a point in which the transformations for the $(\hat u, Y)$ kernels of every mode become singular. 
We explore the implications of this sign change in the next section, and explain its origin in Appendix~\ref{sec:kerprop}.

\section{The Forward Problem}
The ultimate goal of our analysis is to measure the differences in structure between a model and a real star at one or more locations in their interior using the differences in their frequencies. 
This is called the inverse problem, because it is inverse to the forward problem of computing the differences in frequency between such a pair from differences in their internal structure. 
In this section, we investigate the forward problem, and use it to establish the limitations on our ability to solve the inverse problem. 
We carry this out using pairs of models, with one being the reference model and the other serving as a proxy star. 

An assumption underlying a structure inversion is that the kernels accurately translate small differences in internal structure into differences in oscillation mode frequencies. 
The accuracy of the kernels can be assessed by measuring the level of agreement between the left and right sides of Equation~(\ref{eq:inversion}). 
This relation is based on linearization in terms of changes in the frequencies and stellar structure and hence higher-order terms introduce departures from it. 
Also, models that differ in their internal structure may also differ in their kernels, and this introduces errors into the inversion result. 
\citet{2000ApJ...529.1084B} analyzed calibrated solar models and found that the modest differences between the models had little effect on the kernels, causing only small systematic errors in the helioseismic structure inversions. 
In the stellar case, differences between globally fitted models are typically larger. 
Here we assess their effects for subgiants by comparing 
the frequency differences that are computed through Equation~(\ref{eq:inversion}) to the actual frequency differences between various pairs of models. 

%


\subsection{Mode matching} \label{sec:modes}

A basic issue in starting to make such an assessment involves matching up the modes between the pair of models, or between a model and a star. 
On the main sequence, such assignments are achieved in a straightforward manner by either matching radial orders or simply by pairing modes that are closest in frequency. 
Several difficulties arise for subgiant and red giant stars: the bumping of frequencies, which can make pairing based on frequency alone difficult; potentially differing $g$-mode components of the mixed modes owing to their rapid evolution, which can make pairing based on radial order difficult; and the possible emergence of multiple non-radial modes per radial-mode order, the number of which can differ from model to model, thus making the assignment problem unbalanced \citep{2020ApJ...898..127O}. 
These matters are made worse by the surface term, which may cause misidentification of $p$-mode/$g$-mode combinations \citep{2018MNRAS.478.4697B}. 

For this work, we use a simple procedure for mode matching which we have found to work well both for pairs of models and for a model and a star. 
We assume the spherical degrees of the modes have been reliably identified. 
We fit the \citet{2014A&A...568A.123B} two-term surface term correction only to the radial modes, and then apply that correction to all of the modes. 
Finally, we iteratively match together the pairs of modes with the same degree that have the smallest difference in frequency. 

\subsection{Kernel error} \label{sec:ker-errs}
\begin{figure}
    \centering
    \includegraphics[width=\linewidth,trim={0 1.5cm 0 0.6cm}, clip]{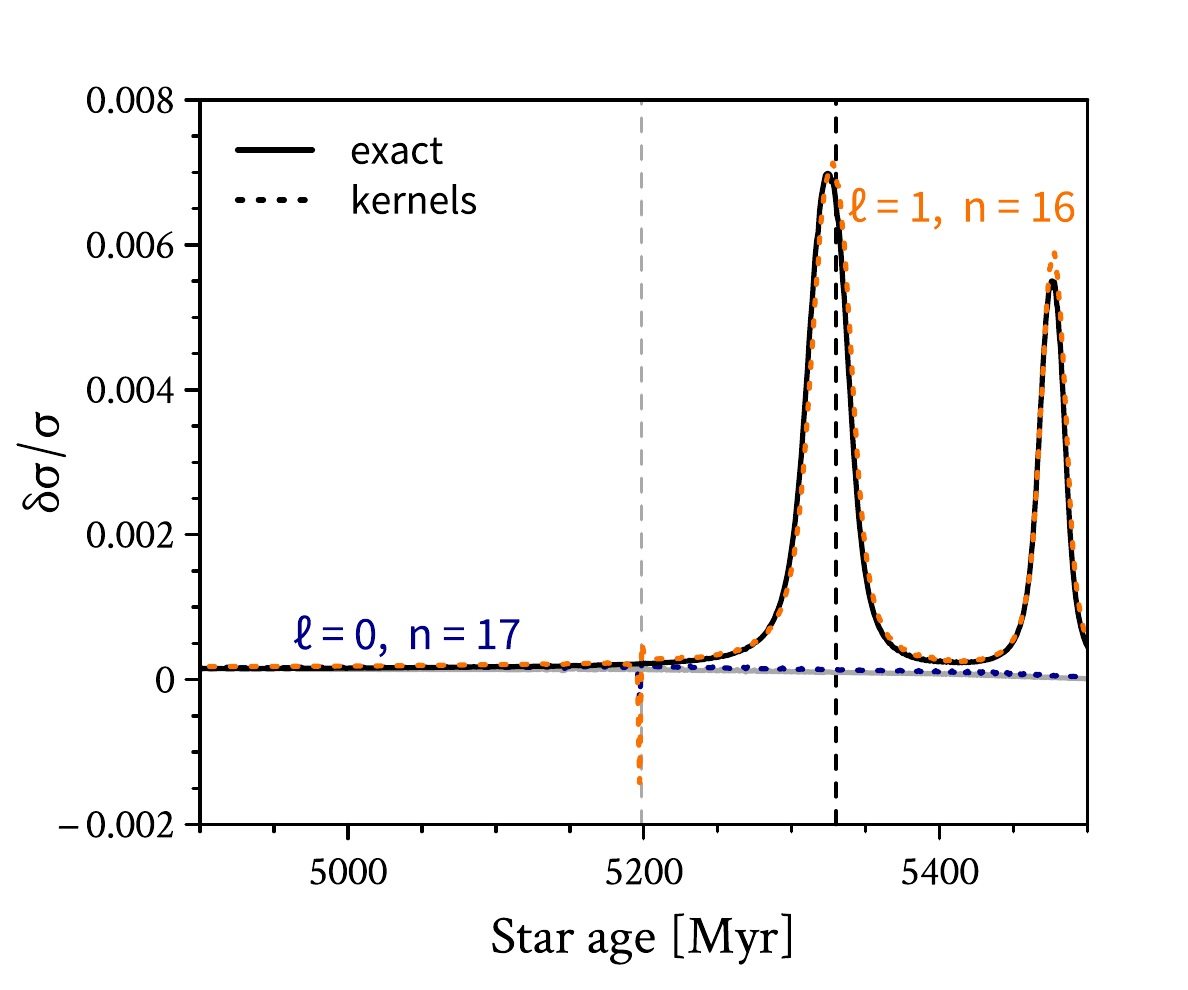}\\%
    \includegraphics[width=\linewidth,trim={0 1.5cm 0 0.6cm}, clip]{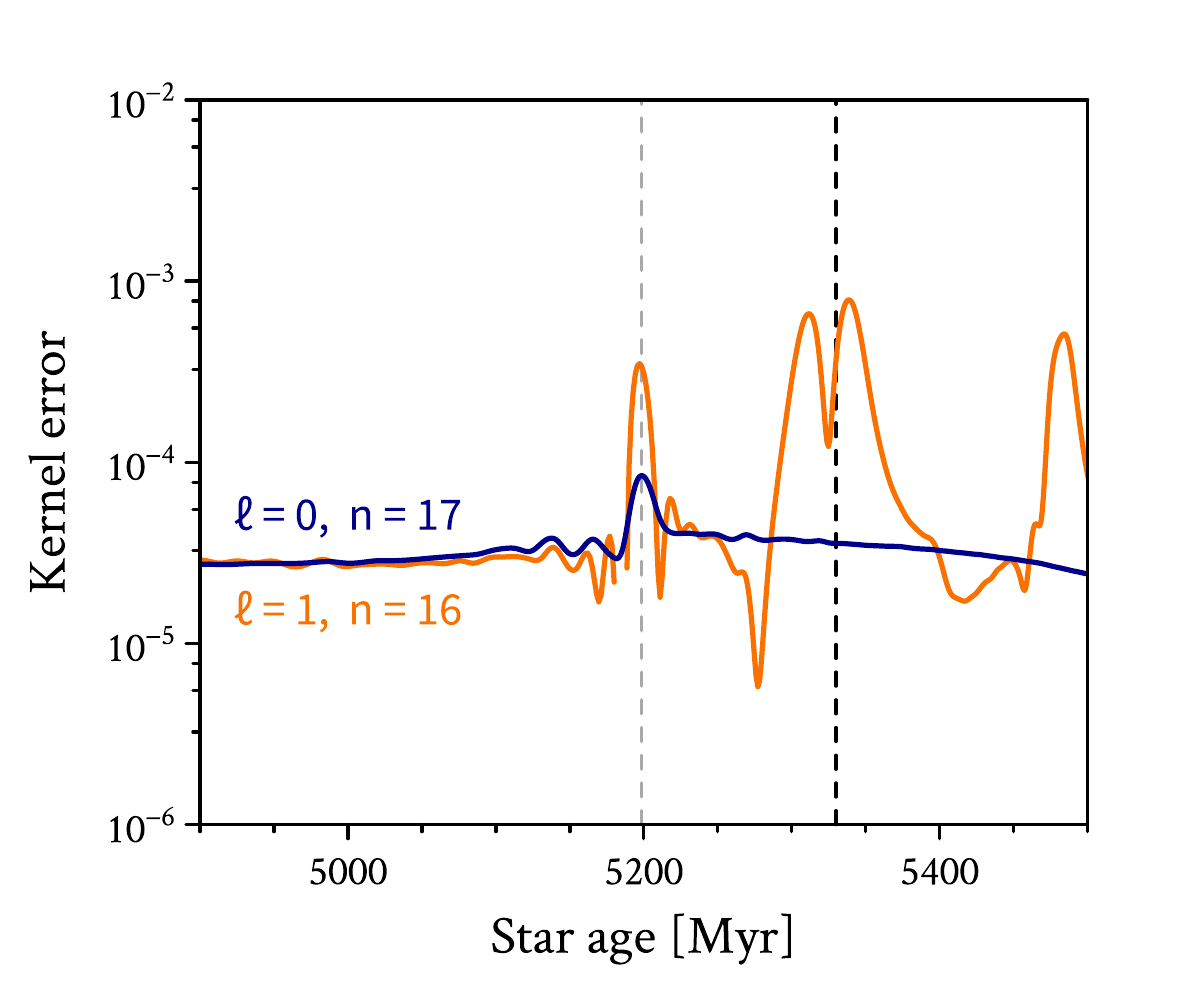}\\%
    \includegraphics[width=\linewidth,trim={0 0 0 0.6cm}, clip]{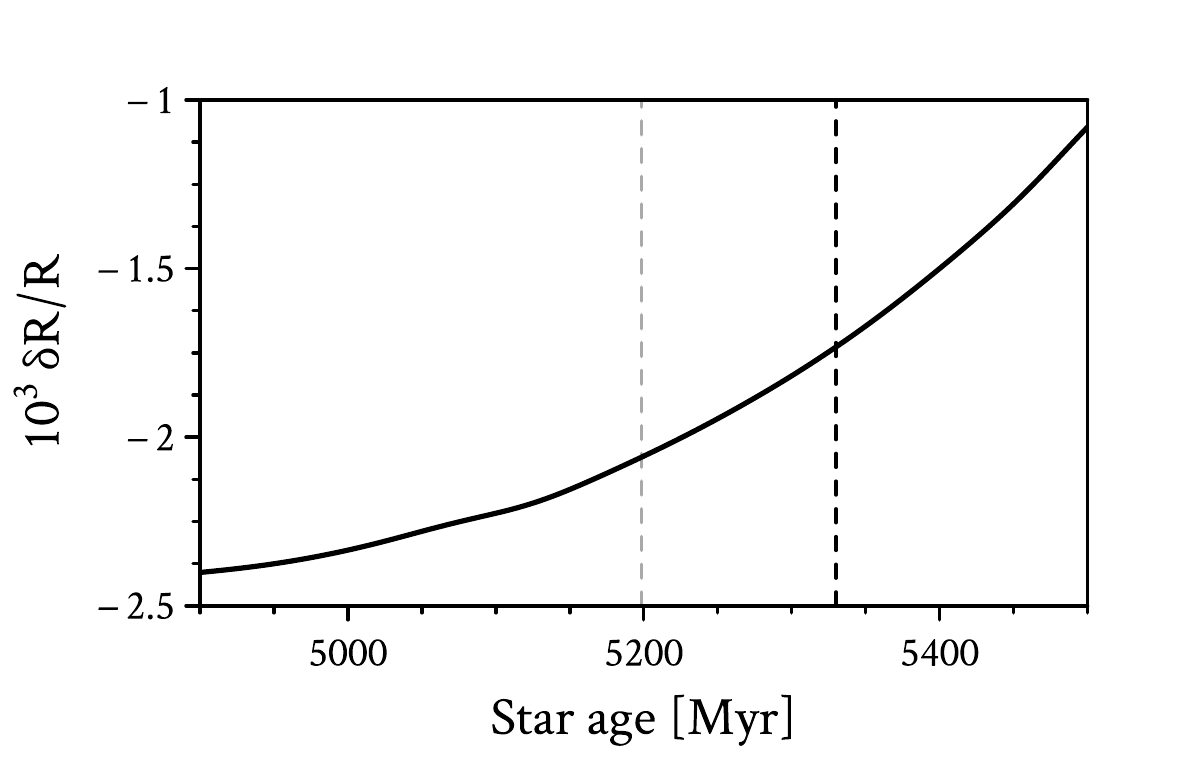}\\%
    \caption{
        \textsc{Top Panel}: \mbb{Frequency differences for one radial (blue/gray) and one dipolar (orange/black) oscillation mode for a $1.2~\text{M}_\odot$ stellar track as a function of age for pairs of models along the same track. 
        The solid black and gray lines give the exact differences, and the dotted blue and orange lines give those calculated using the kernels as in Equation~\ref{eq:inversion}.} 
        The black vertical dashed line indicates the age of the first avoided crossing for the dipolar mode as seen in Figure~\ref{fig:kernel-evol}, and the gray vertical dashed line indicates the age of the sign change seen also in that figure. 
        \textsc{Middle Panel}: Differences between the exact frequency differences and the differences calculated using the kernels, i.e., differences between the solid and dotted lines in the top panel. 
        Smoothing has been applied to eliminate some numerical noise. 
        The kernels are changing most rapidly during the avoided crossings, where the linearization worsens and hence the kernel errors increase. 
        \textsc{Bottom Panel}: Differences in radius. 
        \label{fig:kernel-errs}} 
\end{figure}

Once mode matching has been completed, we may compare the left and right sides of Equation~(\ref{eq:inversion}). 
We call the absolute differences between these two sides the \emph{kernel error}. 
Figure~\ref{fig:kernel-errs} shows the evolution of kernel errors for models along the subgiant track that was shown previously. 
This figure compares each model in the track with the model five time-steps away from it. 
Typical errors are on the order of $10^{-5}$ on the main sequence and along the subgiant branch until the avoided crossings emerge. 
The kernels between the pairs of models differ the most from one another during the avoided crossings, and as a result, the error of the mode undergoing crossing rises to a maximum of nearly $10^{-3}$. 
This figure also demonstrates that the sign change seen in Figure~\ref{fig:kernel-evol} is real, as otherwise the errors would be substantially larger after the sign change. 
This is shown most convincingly by the stability in the errors in the radial mode, for which the kernel also changes sign. 
In Appendix~\ref{sec:kerprop} we furthermore prove the conditions under which the $K^{(\hat u, Y)}$ kernels become singular, thus explaining the sign change.

The models in this test have the same mass and differ in age and radius. 
We will now look at models that differ in mass as well, with differences between the models that span the uncertainties in these parameters.

\subsection{Models of HR~7322}
We have been provided with three best-fitting models by \citet{2019MNRAS.489..928S} from their study of HR~7322, as well as the best-fitting model from \citet{2020MNRAS.499.2445H}.
These models differ in their mass (spanning a range of 
 $0.12~\text{M}_\odot$, or 10\%) 
and radius 
($0.066~\text{R}_\odot$, 3\%) 
as well as in their input physics regarding the treatment of convection, in terms of the mixing length parameter $\alpha_{\text{MLT}}$ as well as the inclusion of overshooting. 
Some parameters of these models are listed in Table~\ref{tab:models}, and the \'{e}chelle diagram for Model~A is shown in Figure~\ref{fig:echelle}. Model~D has an age of 3.2~Gyr, whereas the other models are about 1~Gyr younger.

\begin{table}
    \centering
    \caption{Parameters of the best-fitting models to HR~7322. The first three models are from \citet{2019MNRAS.489..928S}. The last model is from \citet{2020MNRAS.499.2445H}.
    \label{tab:models}}
    \begin{tabular}{ccccc}
        Model & A & B & C & D \\\hline\hline
        $M/\text{M}_\odot$    & 1.198 & 1.249 & 1.239 & 1.314 \\
        $R/\text{R}_\odot$    & 1.951 & 1.961 & 1.969 & 2.017 \\
        $L/\text{L}_\odot$    & 5.377 & 6.054 & 5.970 & 6.070 \\
        $Y_0$    & 0.262 & 0.262 & 0.262 & 0.248 \\
        $M_{\text{core}}/\text{M}_\odot$ & 0.0638 & 0.0680 & 0.0683 & 0.0786 \\
        $\Delta\nu/\mu$Hz     & 54.67 & 55.09 & 54.82 & 54.28 \\
        $\alpha_{\text{MLT}}$ & Subsolar & Solar & Solar & Solar \\
        Overshoot & No & No & Yes & Yes \\\hline
    \end{tabular}
\end{table}

The Stokholm models were generated using the Garching stellar evolution code (\textsc{Garstec}, \citealt{2008Ap&SS.316...99W}) and fit with the Bayesian stellar algorithm (\textsc{Basta}, \citealt{2015MNRAS.452.2127S, 2017ApJ...835..173S}). 
The Hon model is a \textsc{Mesa} model obtained with machine learning, which fit the mass, age, composition, mixing length parameter, overshoot, undershoot, and diffusion factor to the observations. 
Owing to the somewhat larger mass of the star, the fitted diffusion factor is small ($0.06$). 
Notably, the Hon model has a initial helium abundance of $0.248$, which is consistent with the primordial value and thus may be to be too low given that HR~7322 is in the solar neighborhood and has a metallicity of [Fe/H]~$={-0.23 \pm 0.06}$. 

\begin{figure*}
    \centering
    \includegraphics[width=\linewidth,trim={0 0 0 0.5cm}, clip]{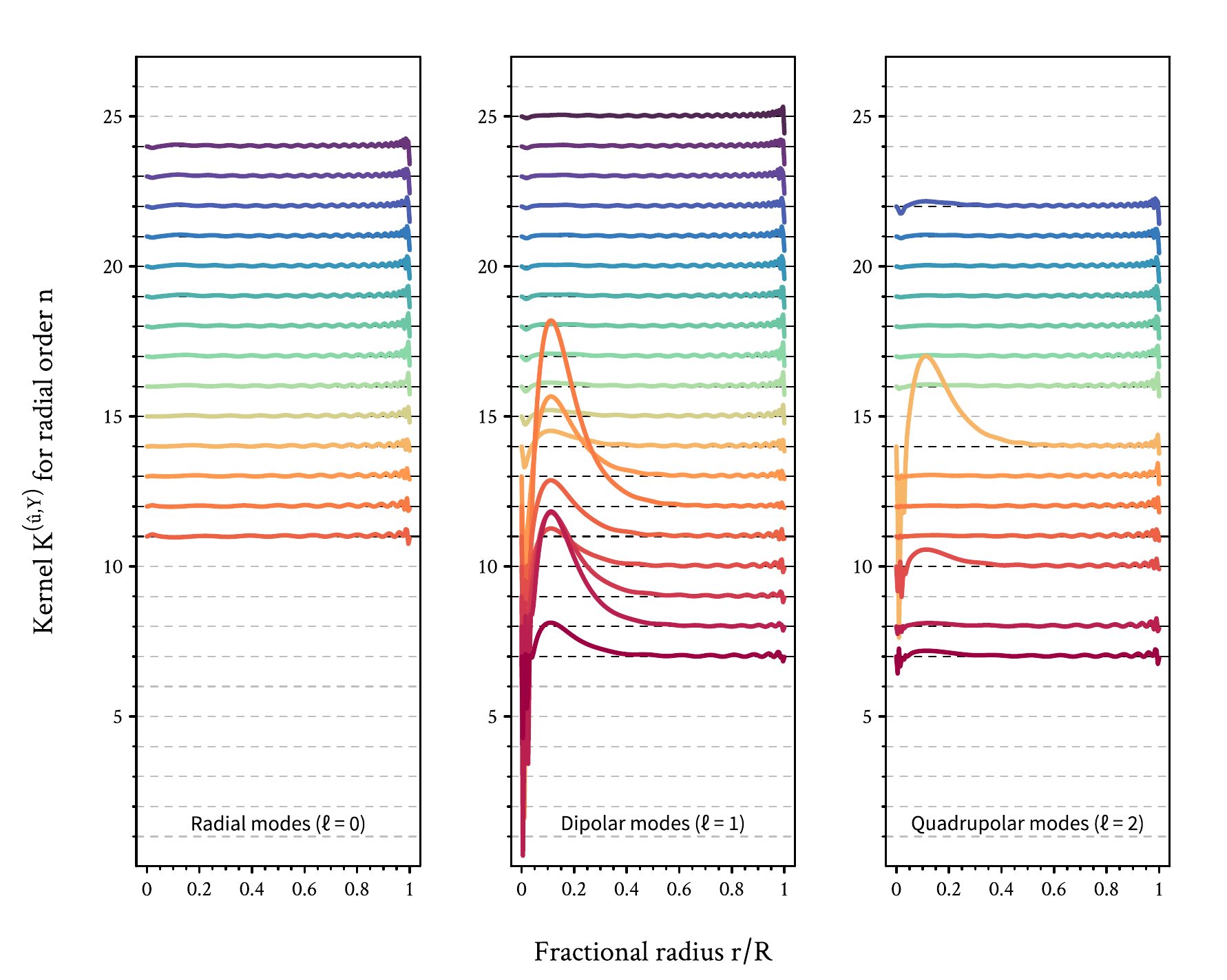}%
    \caption{Structure kernels for Model~A showing the sensitivity of each mode's frequency to a perturbation in $\hat u$. The mode set is the same as was observed in HR~7322. 
    The quadrupolar modes with radial orders $9$ and $15$ were not detected, likely owing to their high inertias. 
    The zeropoint of each kernel is offset by its radial order, and each kernel is normalized by the same fixed value and colored for visibility. 
    \label{fig:kernels}} 
\end{figure*}
\begin{figure*}
    \centering
    \includegraphics[width=\linewidth,trim={0 0 0 0.5cm}, clip]{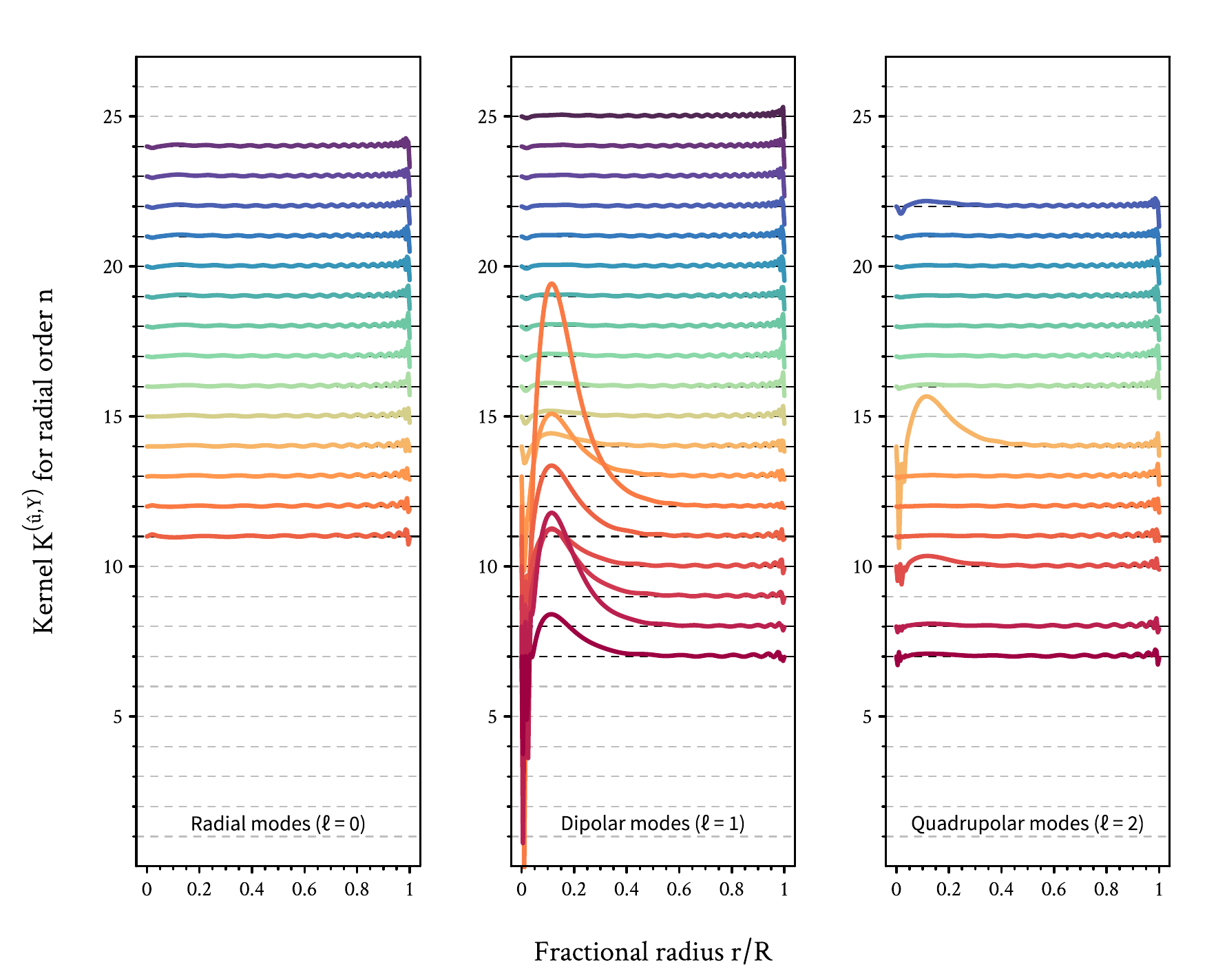}%
    \caption{Structure kernels for Model~B of HR~7322  (\emph{cf}.~Figure~\ref{fig:kernels})
    \label{fig:kernels2}} 
\end{figure*}

A structure inversion assumes that we understand how changes in the structure translate into changes in the oscillation mode frequencies. 
One way to get a sense of how well the kernels will work is by comparing the kernels of different models fit to the same star. 
The structure kernels for Model~A and Model~B are shown in the style of \citet{1994A&AS..107..421A} in Figures~\ref{fig:kernels} and \ref{fig:kernels2}. 
The structure kernels of the two models are similar, differing the most in the amplitude of the $g$-mode components of the mixed modes. 

\begin{figure*}
    \centering
    \vspace*{0.25cm}
    \includegraphics[width=0.85\linewidth,trim={0 0 0 0.5cm}, clip]{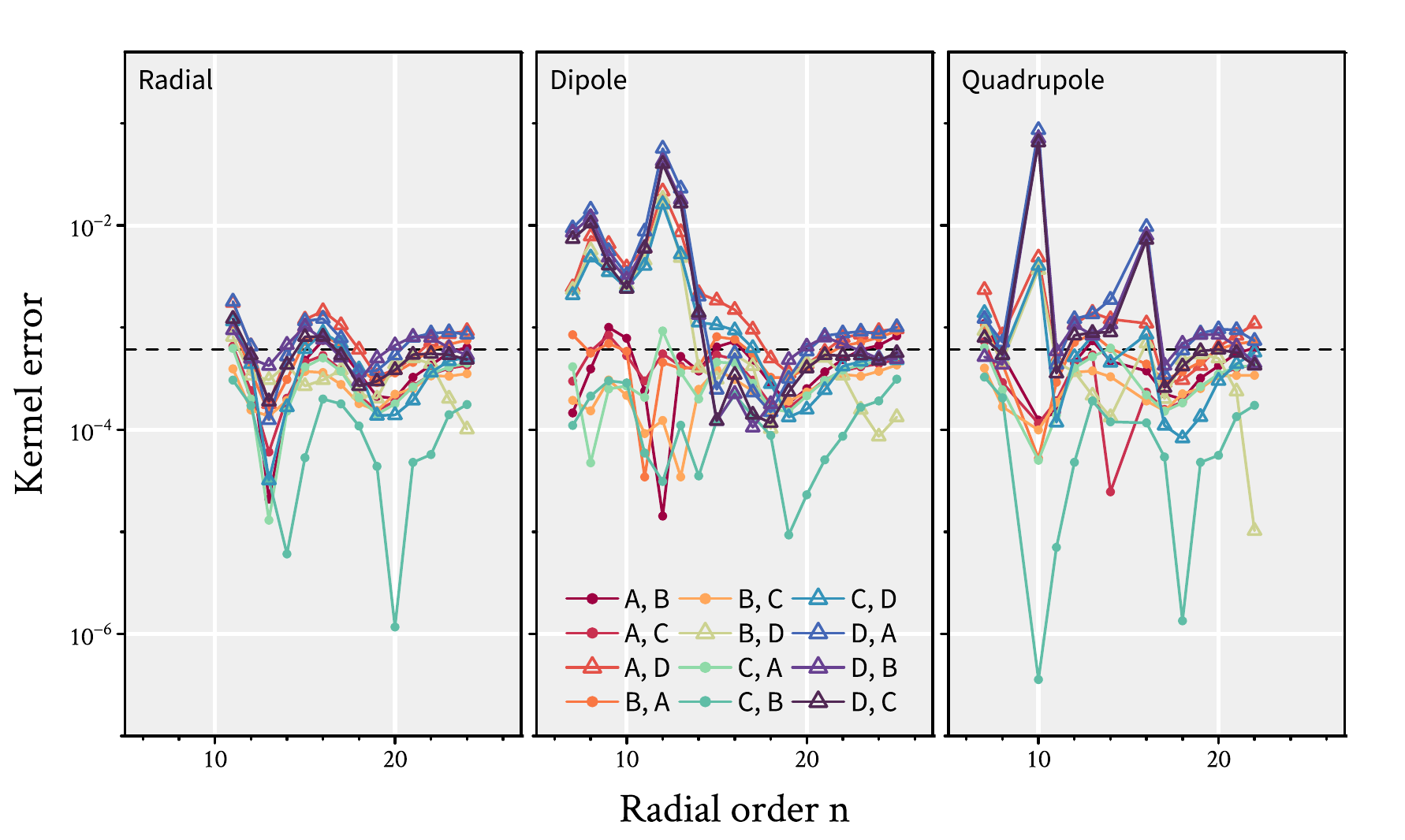}%
    \caption{
        Kernel errors for all pairs of models from Table~\ref{tab:models}. 
        Each color represents a model pair, with the first serving as the reference model and the second as the proxy star. 
        Lines connecting the points are shown to guide the eye. 
        \mbb{The horizontal dashed line indicates the median relative uncertainty of the mode frequencies measured in HR~7322. The open triangles refer to pairs that include Model~D, which show the largest kernel errors as it is the most dissimilar model from the other three.} 
        \label{fig:kernel-errs-unc}}  
\end{figure*}

A comparison of the kernel errors for all pairs of these models is shown in Figure~\ref{fig:kernel-errs-unc} using the mode set observed in HR~7322. 
They span a range of $10^{-6}$ to nearly $10^{-1}$. 
Some of these errors exceed typical relative uncertainties in the observational measurements ($\sim {6\times10^{-4}}$), which should be kept in mind when interpreting inversion results. 
\mbbb{However, as will be seen later, these kernel errors appear to cancel out when performing the inversion.}

\section{The Inverse Problem}
The problem at hand is to infer the internal structure of a star from measurements of its oscillation mode frequencies. 
This problem is more difficult than the forward problem because there are infinitely many possible solutions that are compatible with the oscillation data, including ones which are unphysical. 
Also, small perturbations to the input data may lead to large perturbations in the inversion result (for discussion, see \citealt{1991sia..book..519G} and \citealt{BellingerMJT}). 
In this section, we will again use tests on models to assess our ability to solve the inverse problem. 

\subsection{Optimally Localized Averages}
We employ the method of Subtractive Optimally Localized Averages \citep[SOLA,][]{1992A&A...262L..33P, 1994A&A...281..231P}. 
This method attempts to make a localized estimate of structure at a given target radius $x_0$ by forming an averaging kernel $\mathscr{K}$ which is a linear combination of the mode kernels:
    $\mathscr{K} = \sum_i c_i K_i^{(\hat u,Y)}.$
Provided that $\mathscr{K}$ is unimodular, has the majority of its amplitude at $x_0$ and not elsewhere, and the corresponding cross-term kernel $\mathscr{C} = \sum_i c_i K_i^{(Y,\hat u)}$ is relatively small, then $\delta \hat u/\hat u(x_0)$ is estimated by that linear combination of the oscillation data:
\begin{equation}
    \sum_i c_i \frac{\delta \sigma_i}{\sigma_i} 
    =
    \int \left( \mathscr{K} \frac{\delta \hat u}{\hat u} + \mathscr{C} \delta Y \right) \,\text{d}x
    \simeq 
    \frac{\delta \hat u}{\hat u}(x_0).
\end{equation}
SOLA works by establishing a target kernel, $\mathscr{T}$, and finding the vector $\mathbf c$ that minimizes the residual norm between $\mathscr{T}$ and $\mathscr{K}$. 
Here we use a modified Gaussian function which is peaked at $x_0$ and decays to zero at the center of the star:
\begin{equation} 
    \mathscr{T}(x) = a x \exp\left\{-\left( \frac{x-x_0}{\Delta} + \frac{\Delta}{2x_0} \right)^2 \right\}
\end{equation}
where $a$ is a normalization factor to ensure that $\int \mathscr{T}\,\text{d}x=1$, and $\Delta$ is a free parameter controlling the width of the target kernel. 

The uncertainty $\epsilon$ of the resulting estimate is obtained via the same linear combination applied to the observational uncertainties: $\epsilon^2 = \sum_i c_i^2 s_i^2$, where $s_i$ denotes the uncertainty of $\delta\sigma_i/\sigma_i$. 
Thus, a balance must be struck between forming a well-localized averaging kernel, suppressing the cross-term kernel, and suppressing the amplification of uncertainty. 
These competing objectives can be combined by introducing inversion parameters $\beta$ and $\mu$ and minimizing the functional
\begin{equation} \label{eq:sola}
    \int (\mathscr{K} - \mathscr{T})^2 \,\text{d}x + \beta \int \mathscr{C}^2 \,\text{d}x + \mu \,\epsilon^2
\end{equation}
subject to the constraints that $\int \mathscr{K}\, \text{d}x=1$ and $\sum_i c_i F_{\text{surf}}(\nu_i) = 0$, where here $F_{\text{surf}}$ is the \citet{2014A&A...568A.123B} two-term surface term correction. 
The procedure for minimizing this functional is discussed in Appendix~\ref{sec:minimize}.

\subsection{Nondimensionalization} \label{sec:nondimensionalization}
A basic issue caused by imprecise stellar mass and radius estimates is the nondimensionalization of the mode frequencies that is required for Equation~(\ref{eq:inversion}). 
Such a scaling is trivial for pairs of models, as their mass and radius are known exactly. 
When analyzing observed frequencies, and hence {\it a priori} unknown mass and radius, there are two approaches for dealing with this issue \citep{Basu2003}. 
The first is to include and minimize an additional term $\frac{1}{2}\delta \ln (M/R^3)$ and corresponding free parameter to Equation~(\ref{eq:sola}) \citep{1993ASPC...40..541G, 2020IAUS..354..107K}. 
The second approach, which we take in this work, is to subtract from all the modes a weighted mean of the differences in radial mode frequencies, which emulates the scaling \citep{1998IAUS..181P....P}.

\subsection{Model Test Results}
\begin{figure}
    \centering
    \vspace*{1cm}
    \includegraphics[width=\linewidth,trim={0 0 0 0.5cm}, clip]{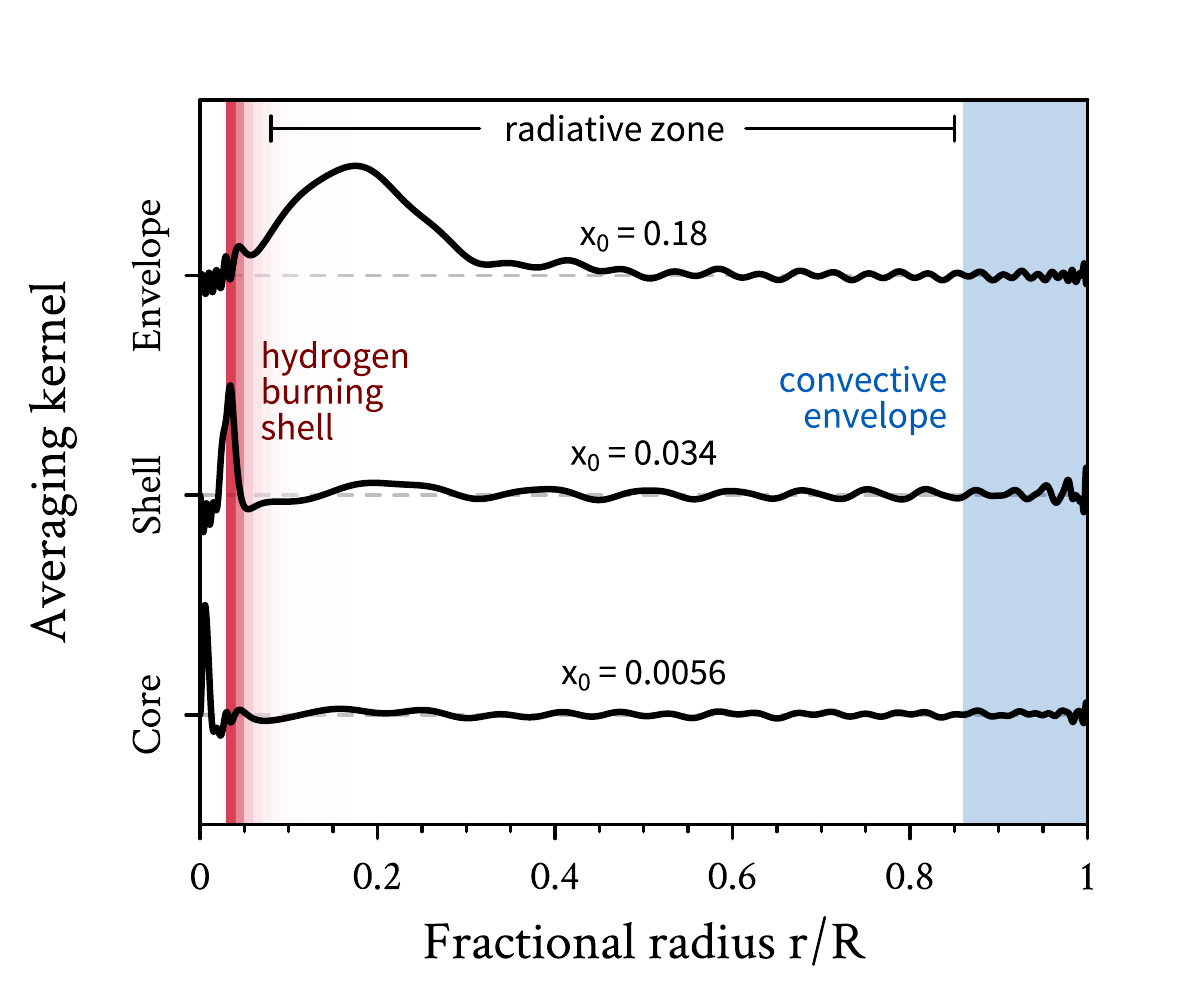}%
    \caption{Averaging kernels $\mathscr{K}^{(\hat u,Y)}$ for Model~A that are localized at three different radii as indicated. Each averaging kernel is normalized by its maximum amplitude and offset from one another for the sake of visibility. Some aspects of the reference model are labeled. The intensity of the red color is proportional to the energy generation rate per unit mass in the reference model, up to a maximum of 138~erg~s$^{-1}$~g$^{-1}$. 
    \label{fig:avgk}} 
\end{figure}

\begin{figure}
    \centering
    \vspace*{1cm}
    \includegraphics[width=\linewidth,trim={0 0 0 0.5cm}, clip]{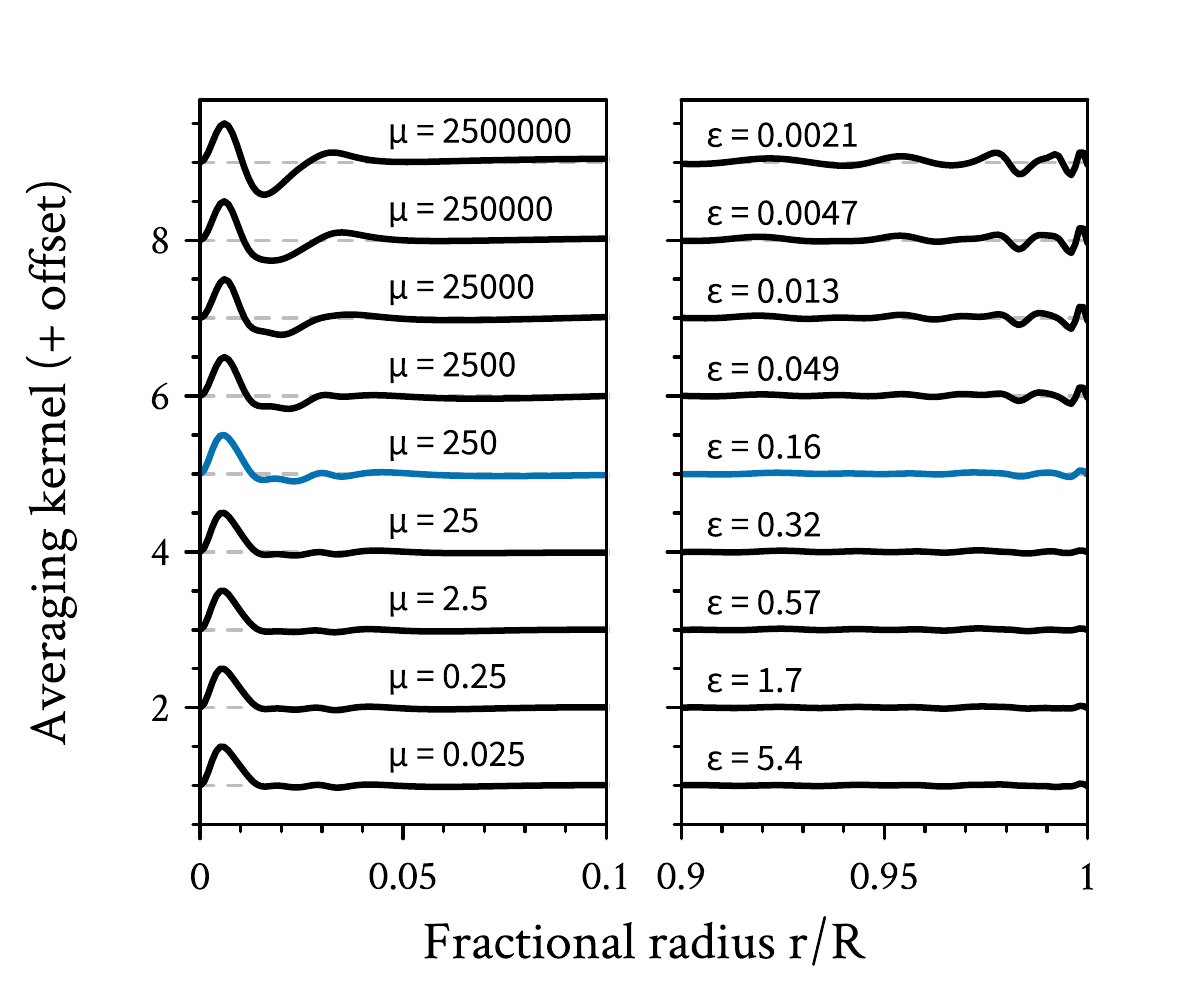}%
    \caption{Core averaging kernel for Model~A for different choices of the error suppression parameter $\mu$. 
    The uncertainty $\epsilon$ associated with each choice of $\mu$ is indicated. 
    The left panel shows the behavior in the core and the right panel shows the behavior near to the surface. 
    The middle averaging kernel shown in blue is the same as in Figure~\ref{fig:avgk}.
    Large values of $\mu$ yield averaging kernels with substantial sidelobes and large surface features, and hence have no resolving power; and low values of $\mu$ yield perfectly localized kernels, but with undesirably large uncertainty. \label{fig:mu} } 
\end{figure}

%
By manually selecting the parameters $x_0, \Delta, \mu,$ and $\beta$ (see Table~\ref{tab:inv-params}) and inspecting the result, we have been able to form three well-localized averaging kernels for HR~7322, which are visualized in Figure~\ref{fig:avgk} along with some aspects of the reference model. 
They are positioned in the inert helium core (${x_0 = 0.0056}$), hydrogen burning shell (${x_0 = 0.034}$), and deeper part of the radiative envelope of the star (${x_0 = 0.18}$). 
For these averaging kernels, we found it unnecessary to suppress the cross-term kernel, as it has negligible amplitude regardless of the value of $\beta$, and so we use ${\beta=0}$ throughout. 
To give some insight into the effect of the inversion parameters, Figure~\ref{fig:mu} shows how the core averaging kernel changes based on different choices of the error suppression parameter $\mu$. 

\begin{table}
    \centering
    \caption{Inversion parameters for HR~7322 corresponding to the three averaging kernels shown in Figure~\ref{fig:avgk} \label{tab:inv-params}}
    \begin{tabular}{c|cccc|c}
        & $x_0$ & $\Delta$ & $\mu$ & $\beta$ & $\epsilon$ \\\hline\hline
        Core & $0.0056$ & $0.0056$ & $250$ & $0$ & $0.16$ \\
        Shell & $0.034$ & $0.011$ & $1000$ & $0$ & $0.047$ \\
        Envelope & $0.18$ & $0.11$ & $6.3$ & $0$ & $0.049$ \\\hline
    \end{tabular}
\end{table}

It is worth mentioning that thanks to the mixed oscillation modes, the innermost averaging kernel is localized deeper within this star than that which is possible with present helioseismic data. 
However, it does come with a somewhat substantial uncertainty of $0.16$, which is approximately three times larger than the uncertainties of the other two averaging kernels. 
These averaging kernels are also generally deeper within the star than averaging kernels for other main-sequence stars, which can only be localized in the region $x_0 = 0.05 - 0.35$ \citep{1993ASPC...40..541G, 2003Ap&SS.284..153B, 2017ApJ...851...80B, 2019ApJ...885..143B}. 

\begin{figure}
    \centering
    \includegraphics[width=\linewidth,trim={0 0 0 0.5cm}, clip]{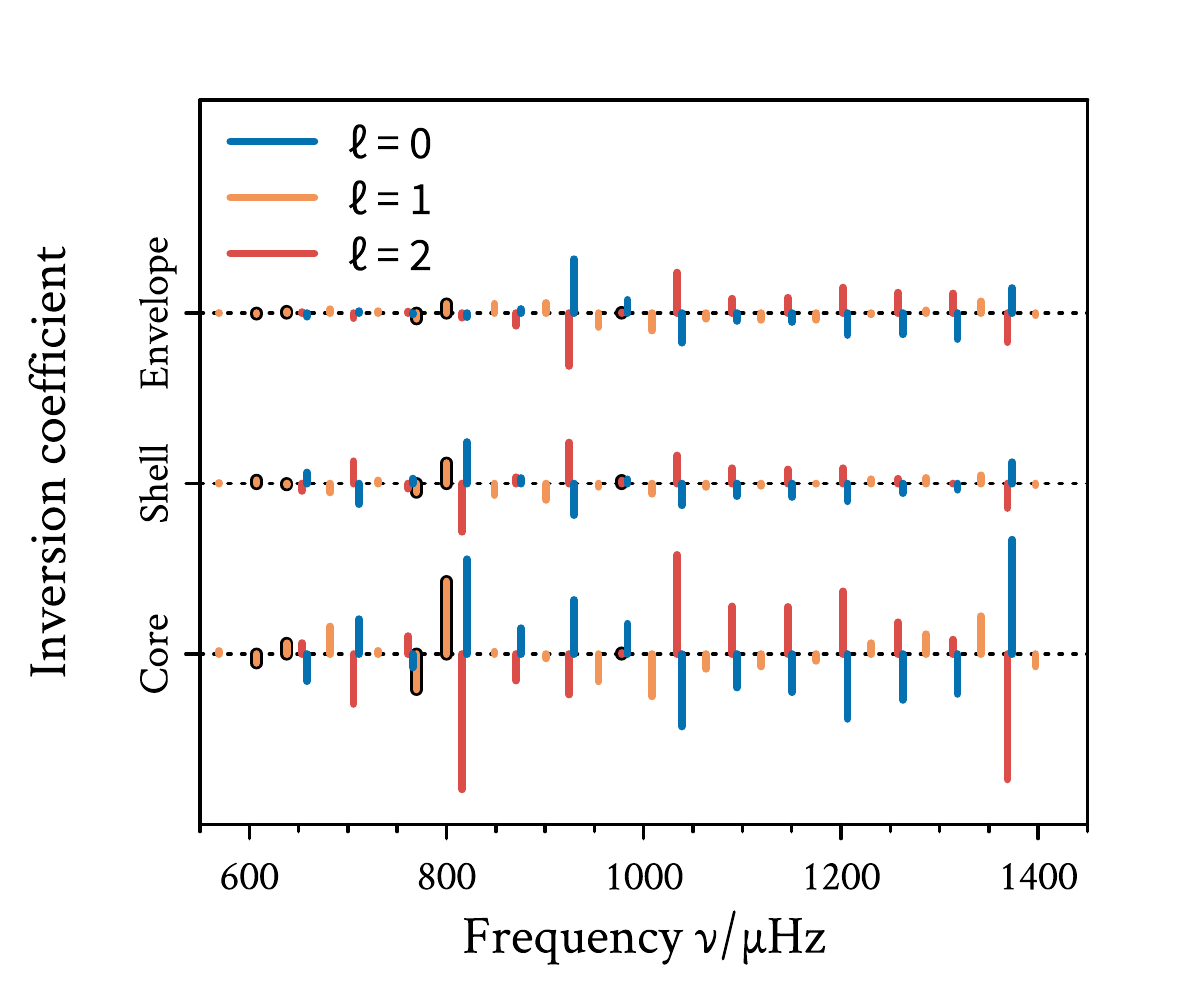}
    \caption{\mb{Inversion coefficients $\mathbf c$ for the core, shell, and envelope averaging kernels shown in Figure~\ref{fig:avgk}. In order to highlight the mixed modes, the non-radial modes are outlined in black if the inertia of the mode exceeds the radial-mode inertia interpolated to that mode's frequency by at least 20\%. }
    \label{fig:inv_coefs}}
\end{figure}

\mb{The inversion coefficients $\mathbf c$ used to construct these averaging kernels are shown in Figure~\ref{fig:inv_coefs}. A few aspects are notable. 
Firstly, as expected, the core averaging kernel makes the most use of the mixed modes, which is evident by their large inversion coefficients. On the other hand, the envelope averaging kernel suppresses the mixed modes. 
Second, the small frequency separation---i.e., the differences between neighboring quadrupolar and radial oscillation frequencies---is evident in all of the averaging kernels, which can be seen through the nearly equal and opposite inversion coefficients for nearly all pairs of $\ell=2$ and $\ell=0$ modes.
This is unsurprising, as it is well-known that the small frequency separation is useful for probing the conditions in the inner layers of a star \citep{1984srps.conf...11C, 2003Ap&SS.284..165G}. 
Finally, it is notable that, relative to the other kernels, the core averaging kernel has more extreme values for the inversion coefficients, resulting in a larger standard deviation of the inferred difference. 
}

\begin{figure}
    \centering
    \includegraphics[width=\linewidth,trim={0 0 0 0.5cm}, clip]{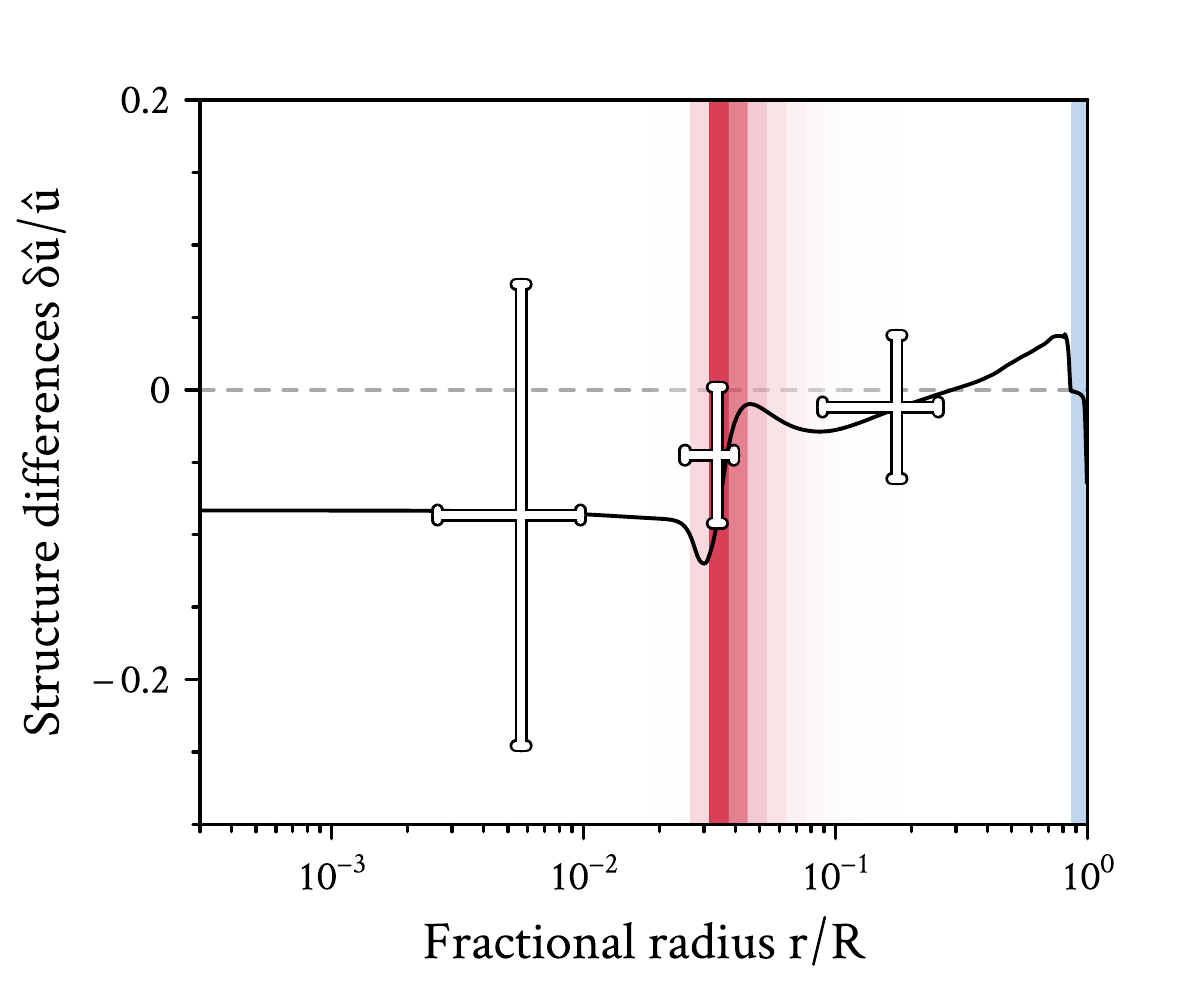}%
    \caption{Model test inversion results where Model~A is chosen as the reference model and Model~D acts as the proxy star. The actual differences in structure are given by the solid black line, and the points give the inversion result of inferring the structure from the seismic data. A negative number indicates that the proxy star has a lower $\hat u$ than the reference model. The vertical error bars indicate the uncertainty in the inversion result, and the horizontal bars depict the full width at half maximum of the corresponding averaging kernel. The background features are the same as in Figure~\ref{fig:avgk}. 
    \label{fig:model-inv}} 
\end{figure}

With such averaging kernels in hand, we may now assess the ability to infer each model's structure from its oscillation data using each other model as the reference. 
Figure~\ref{fig:model-inv} shows the result of inferring the structure of Model~D using Model~A as reference. 
There it can be seen that the differences in structure are recovered well. \mbbb{These results are in spite of the substantial kernel errors seen previously involving Model~D. 
This can be understood by applying the same linear combination used to form the averaging kernels to the kernel errors. 
For these models, these combinations of the kernel errors turn out to be always smaller---and on average about one order of magnitude smaller---than the uncertainty $\epsilon$ associated with each of the averaging kernels. Therefore, the kernel errors cancel out for these models. \mbbbb{Such a cancellation might not happen in all cases.} } 

From inspecting all other pairs of models, we find that the outer kernel performs the best, with a median true error in the inversion result of $0.0019$. The core averaging kernel comes in second with a median true error of $0.010$, and the shell error is last with $0.032$. 
This is not surprising as the sound speed variations between the models change more rapidly in the shell than the resolution of the averaging kernel. 
Nevertheless, these errors are smaller than the uncertainties associated with the corresponding averaging kernels (0.16, 0.047 and 0.049 for the core, shell, and envelope, respectively). 

\mbbb{As discussed in Section~\ref{sec:modes}, a potential issue is in the identification of the modes, as a misidentified mode would cause the wrong kernel to be used in the inversion. 
These four models tend to agree on the identification of the modes except in the case of the ${\ell=2}$ mode at $760.2\,\mu$Hz, in which the Model~D finds a closer match with the ${n=9}$ mode where the others have ${n=10}$. We have performed the inversions with both candidate mode identifications, and found that it made essentially no difference to the results, due to this mode having small inversion coefficients (\emph{cf.}~Figure~\ref{fig:inv_coefs}).
} 

\begin{figure}
    \centering
    \includegraphics[width=\linewidth, trim={0 1cm 0 0.5cm}, clip]{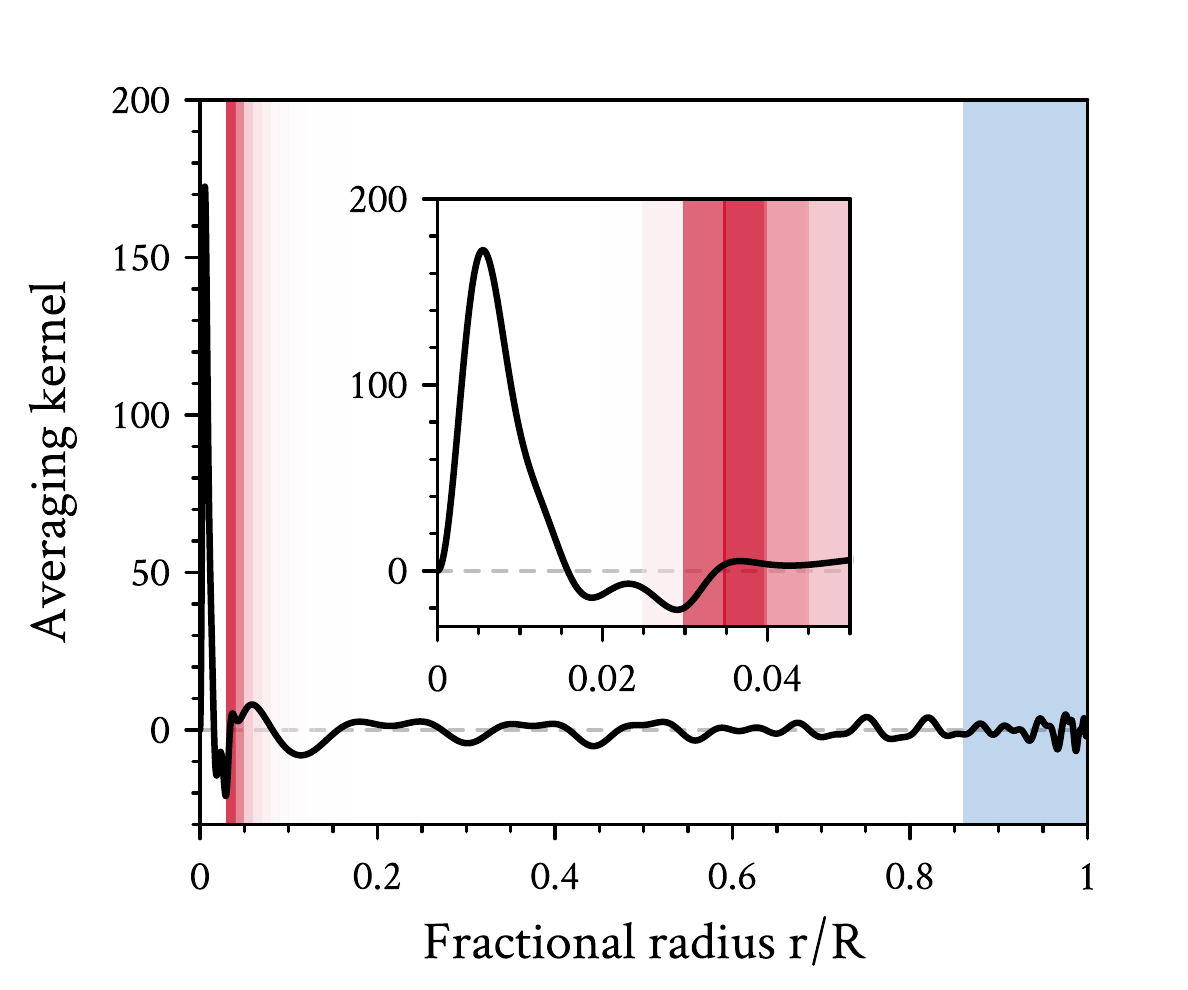}\\%
    \includegraphics[width=\linewidth, trim={0 0 0 0.6cm}, clip]{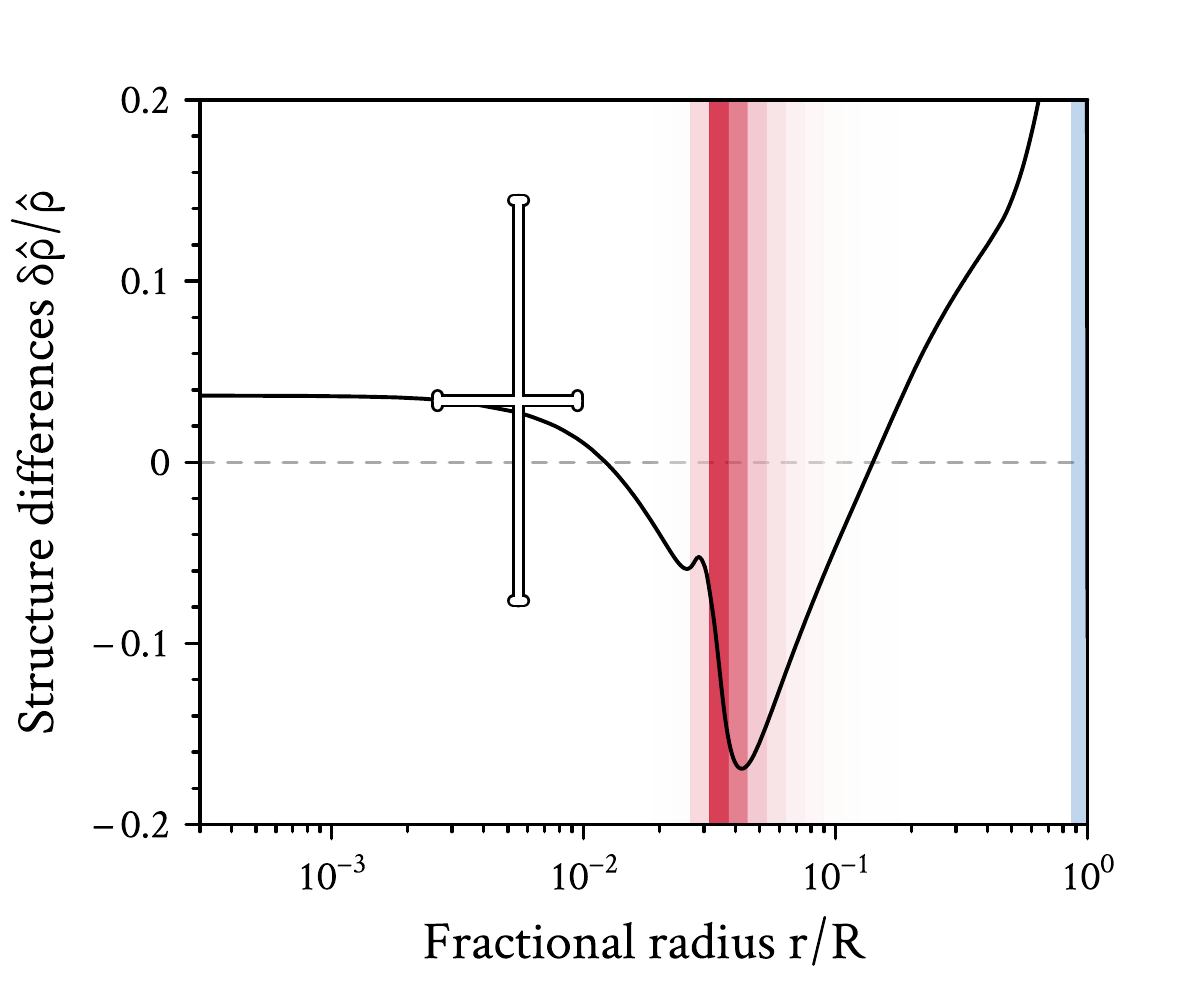}
    \caption{\mb{\textsc{Top Panel}: Averaging kernel $\mathscr{K}^{(\hat \rho, Y)}$ for Model~A that is localized near to the center (\emph{cf.}~Figure~\ref{fig:avgk}). The inset shows a zoom into the core. \textsc{Bottom Panel}: Model test inversion results between Model~A and Model~D (\emph{cf.}~Figure~\ref{fig:model-inv}). Note the difference in the x-axis scale. }
    \label{fig:rho}}
\end{figure}

\mb{Finally, we have also investigated the $(\hat\rho,Y)$ kernel pair for these models. Unlike the work of \citet{2020IAUS..354..107K}, we are only able to form a well-localized $(\hat\rho,Y)$ averaging kernel in the core, and not elsewhere. 
\mbb{This is likely due to the evolving nature of the mixed modes, which for each subgiant are sensitive to different regions of the interior (\emph{cf.}~Figure~\ref{fig:kernel-evol}).}
For the core averaging kernel, we found that the same inversion parameters as in the previous case worked well. The averaging kernel and the model test results between Model~A and Model~D are shown in Figure~\ref{fig:rho}. }

\needspace{3\baselineskip}
\section{Results \& Conclusions}
We now apply the inversion method of the previous section to HR~7322. 
As we have used the mode set and uncertainties from HR~7322 to form the averaging kernels in the previous section, we may use the same averaging kernels to infer the internal structure of HR~7322, and obtain the inversion result by using the observed frequencies rather than the model frequencies. 
This is why it is commonly stated that SOLA-based structure inversions do not fit for the internal structure. 
Rather, SOLA inversions develop an instrument---the averaging kernel---which can then be applied to observations to reveal the internal stellar structure. 

\begin{figure}
    \centering
    \includegraphics[width=\linewidth,trim={0 1cm 0 0.5cm}, clip]{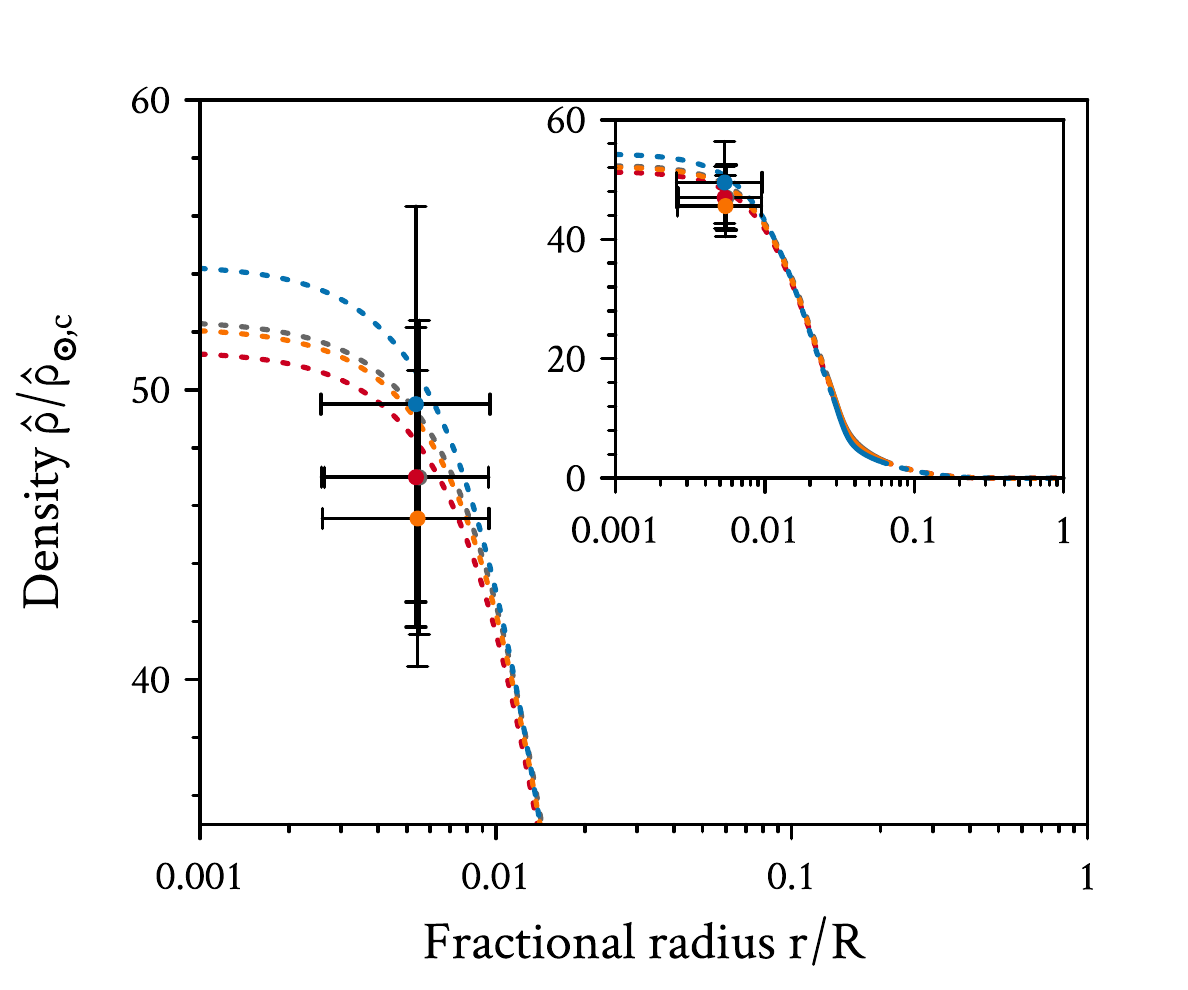}\\
    \includegraphics[width=\linewidth,trim={0 0 0 0.6cm}, clip]{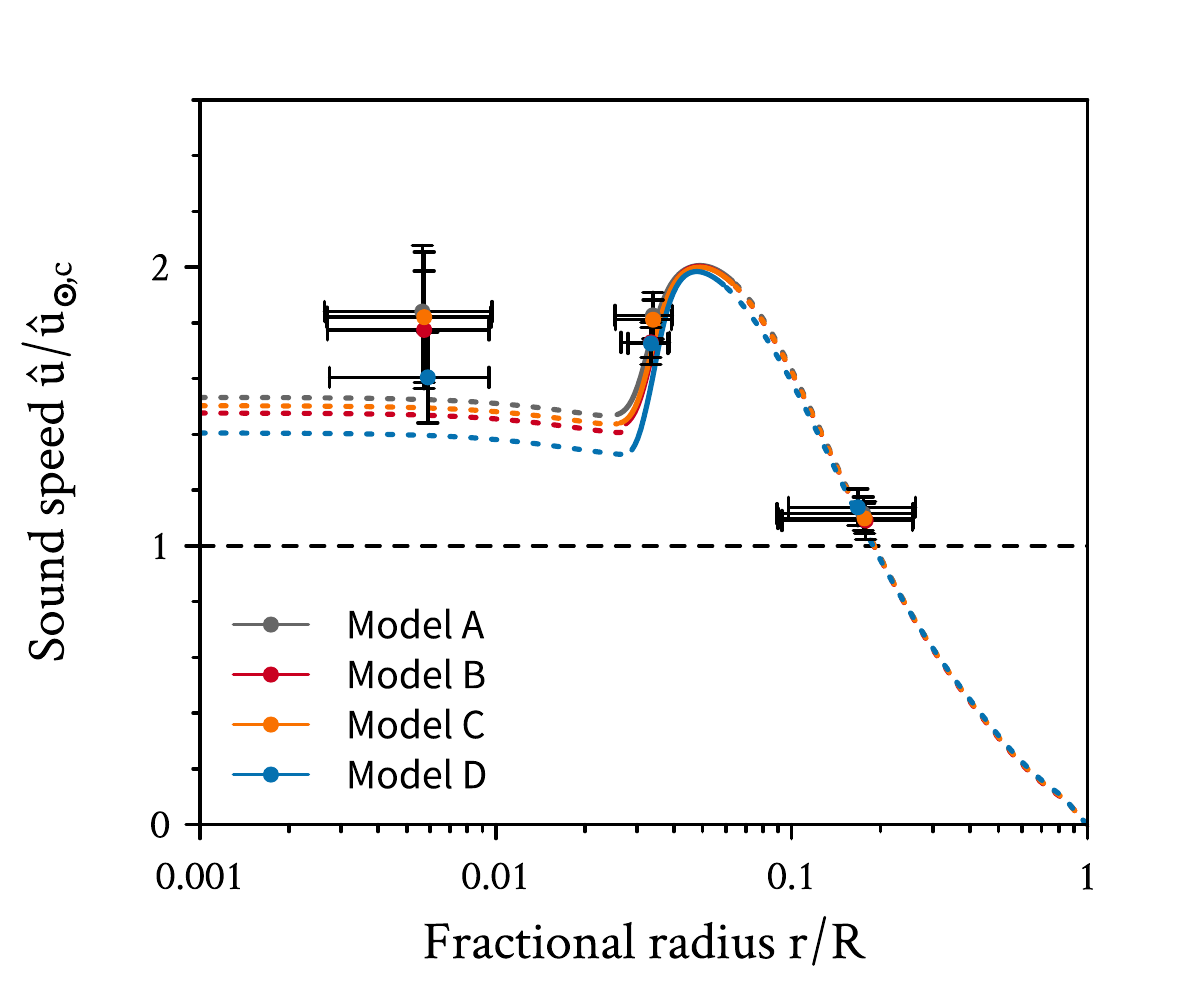}%
    \caption{
    \mb{Inversion results showing the dimensionless density ($\hat \rho$, top panel) and dimensionless squared isothermal sound speed ($\hat u$, bottom panel) measured in HR~7322 (points with error bars) compared with four stellar evolution models (lines). 
    Each of the points shows the result of performing the inversion with respect to the indicated reference model. 
    Density inversions were possible in the inert helium core, whereas the sound speed could be resolved in the core, the hydrogen burning shell, and in the radiative envelope. 
    As in Figure~\ref{fig:soundspeed}, solid lines indicate regions of nuclear fusion, and the $\odot,c$ subscript indicates the corresponding central solar value.} 
    \label{fig:inv}} 
\end{figure}

The results of the inversion for each of the evolutionary models are shown in Figure~\ref{fig:inv}. 
\mbb{The stellar measurements were obtained by correcting the reference model using the inversion result $\delta \hat u/\hat u$ via 
${
    \hat u_{\text{star}} 
    = 
    \left(\delta \hat u/ \hat u + 1\right) \hat u_{\text{model}}
}$, 
and likewise for the density.}
As before, we are able to infer the relative differences in $\hat u$ between the models and the star within the inert helium core, the hydrogen burning shell, and the deeper part of the radiative envelope\mb{, and the relative differences in $\hat \rho$ in the core}. 
\mbb{A clear spread between the different reference models is visible in the inversion results for the core, which arise in connection with the kernel errors seen in Section~\ref{sec:ker-errs}. That being said, the results for the different reference models are consistent within the error bars determined by the statistical errors.} 

\mbb{Whereas all of the models have core densities that are consistent within uncertainty with that of the star, 
we find that the sound speeds in the cores of the models are too slow by a bit more than one standard deviation across every model.} 
As the models have helium cores, this result implies that the cores of the models may be too cool. 
Modeling has shown that the temperature of the core on the subgiant branch can be increased by including the effects of element diffusion. 
However, diffusion in models with masses around and above $1.2~\text{M}_\odot$ such as these is well-known to cause the surface metallicity to drop below observed values. 
This may imply the need to properly account for diffusion and any effects that resist it when modeling solar-like oscillators, such as radiative acceleration \citep[e.g.,][]{2018A&A...618A..10D}, but this will require further study.
These results are similar to those of the main-sequence star 16~Cyg~A, in which the core sound speed of the model seems to be too low \citep{2017ApJ...851...80B}. 
Models of KIC~6225718 also appear to have issues in the core, although in that case the core sound speed is too high \citep{2019ApJ...885..143B}.

\mb{It is notable that, relative to its own inversion result, Model~D appears to perform better in the core than the first three models.} 
However, the differences are not significant: the uncertainty of the inversion result is larger than the differences between the models, and so it is not possible with current data to tell which input physics to evolutionary models are favored in subgiant models.

\begin{figure}
    \centering
    \includegraphics[width=\linewidth,trim={0 0 0 0.6cm}, clip]{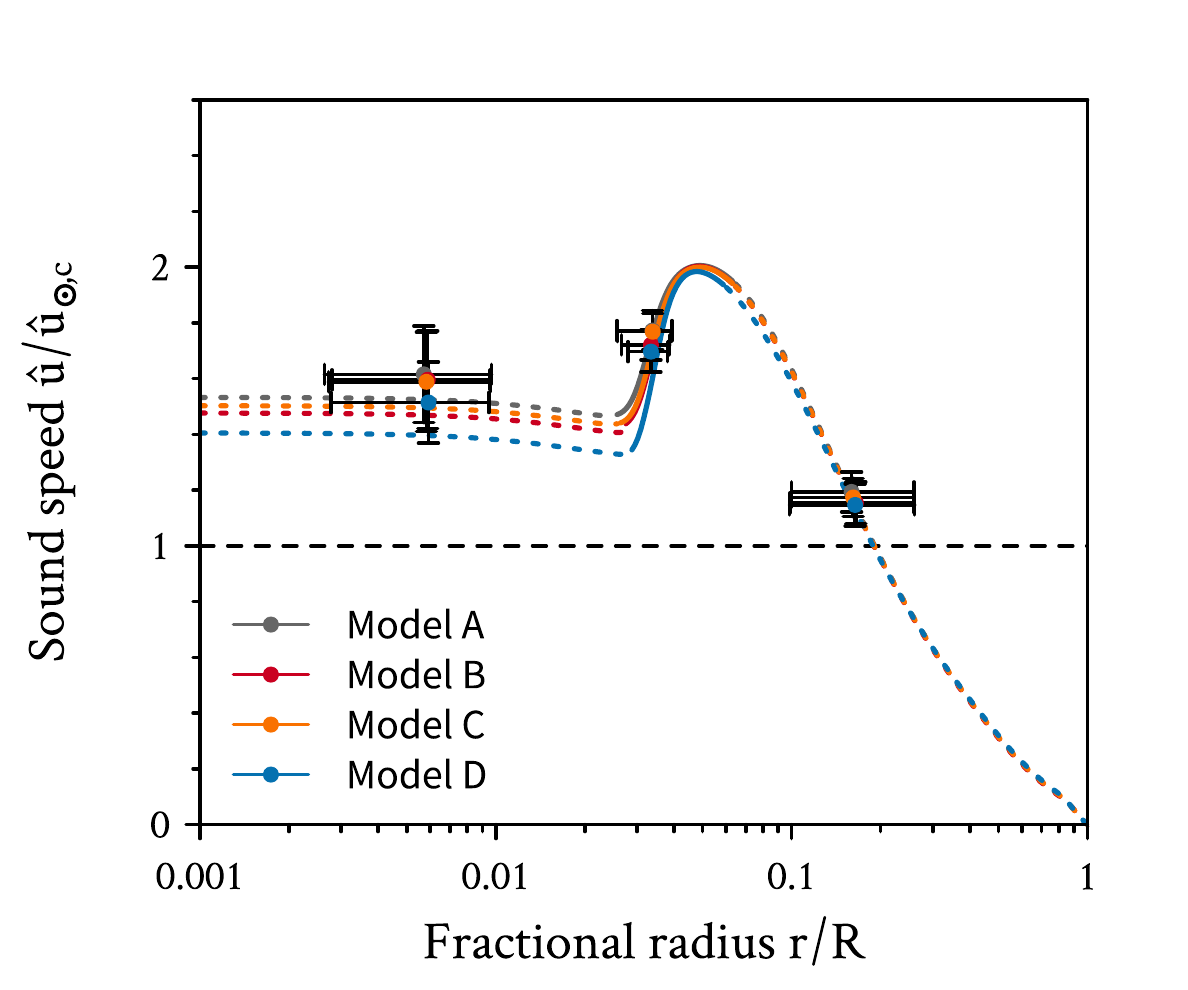}%
    \caption{
    \mbbb{The same as the lower panel of Figure~\ref{fig:inv}, but with the highest-frequency ${\ell=2}$ mode removed.} 
    \label{fig:l2removed}} 
\end{figure}

\mbbb{Figure~\ref{fig:inv_coefs} showed that the core averaging kernel has a large dependence on the highest frequency ${\ell = 2}$ mode. 
An inspection of the \'echelle diagram (Figure~\ref{fig:echelle}) indicates however that this mode could be unreliable, as it does not match the pattern of the other ${\ell=2}$ modes well. 
This mode's frequency also differs from that of each of the models by between $2.4$ and $3.4$ standard deviations, which is generally worse than the other ${\ell=2}$ modes. 
When we repeat the inversion with this mode removed, the core sound speed matches within one standard deviation all of the reference models (see Figure~\ref{fig:l2removed}). Furthermore, with this mode removed, the agreement in the inversion results across the four models improves. 
As expected from Figure~\ref{fig:inv_coefs}, the results in the shell and envelope are essentially unchanged. The same is true for the core density inversion, the estimate of which across the four reference models remains within one standard deviation uncertainty after this mode has been removed. }

In a future work, we will apply these methods to more subgiant stars, and determine whether it will be possible with more data to distinguish between evolutionary models constructed with differing input physics. 
Despite the uncertainties associated with the analysis, it may be possible with a sample of stars to find a consistent result---such as core temperatures that are systematically too low---which would then aid in guiding improvements to stellar evolutionary modeling. 

\acknowledgments{
We thank A.~Stokholm for kindly providing models of HR~7322, and the anonymous referee for the very useful comments that have improved this paper.
Funding for the Stellar Astrophysics Centre is provided by The Danish National Research Foundation (Grant agreement no.: DNRF106). 
}

\software{\textsc{Adipls} \citep{2008Ap&SS.316..113C}, \textsc{Cholmod} \citep{10.1145/1391989.1391995}, \textsc{Mesa} \citep{2011ApJS..192....3P,2013ApJS..208....4P,2015ApJS..220...15P,2018ApJS..234...34P, 2019ApJS..243...10P}, \textsc{magicaxis} \citep{magicaxis}, R \citep{r}}

\needspace{\baselineskip}%
\appendix%
\section{Properties of the kernels}%
\label{sec:kerprop}
In the analysis we take out the dependence on stellar mass $M$ and radius $R$
by working in terms of scaled variables:
\begin{align}
    x &= r/R 
    \; , \quad 
    &&\pp = \left( {G M^2 \over R^4} \right)^{-1} P 
    \; , \\ 
    \rrho &= \left( {M \over R^3} \right)^{-1} \rho 
    \; , \quad
	&&\uu = \left( {M \over R} \right)^{-1} {P \over \rho}
\end{align}
where $G$ is the gravitational constant,
and the dimensionless frequency $\sigma$ as introduced in Equation~(\ref{eq:sigma}).
This leads to Equation~(\ref{eq:inversion}) for the relation between the dimensionless
frequency differences and the model differences.
For tests of the inversion on model differences this scaling is trivial; 
in analysis of observations, we use a procedure developed by
\citet{1998IAUS..181P....P}, mentioned in Section~\ref{sec:nondimensionalization}.

The derivation of the kernels $K^{(\uu, Y)}$ and $K^{(Y, \uu)}$ starts from
the kernels for the variable pair $(\Gamma_1, \rrho)$,
with $\Gamma_1 = (\partial \ln P/\partial \ln \rho)_{\rm ad}$ 
being the first adiabatic exponent, 
which are relatively straightforward to derive \citep{1991sia..book..519G}.
We base the analysis on the Appendix of \citet{2002ESASP.485...95T}.
To obtain the kernel we need to transform integrals involving $\delta \hat \rho$ to integrals involving $\delta \uu$. 
This is accomplished by introducing an auxiliary function $\psi$ such that 
\begin{equation}
    \int_0^1 F\, {\delta \rrho \over \rrho}\, \dd x = 
    - \int_0^1 \pp \left( {\psi \over \pp} \right)' {\delta \uu \over \uu}\, 
	\dd x
\end{equation}
for any function $F$;
here the prime indicates differentiation with respect to $x$.
This is satisfied if $\psi$ satisfies the differential equation%
\footnote{In the corresponding equation on dimensional form the term in $\psi$
in addition includes the factor $G$.}
\begin{equation} \label{eq:psieq}
    \left( {\psi' \over x^2 \rrho} \right)' + {4 \pi \rrho \over x^2 \pp} \psi =
    \left( { F \over x^2 \rrho } \right)'
\end{equation}
with the boundary conditions $\psi(0) = \psi(1) = 0$. 
In the case of obtaining kernels for $\uu$ and $Y$, we have
\begin{equation}
    F = \left[
        \left(
            \frac{\partial \ln \Gamma_1}{\partial \ln P}
        \right)_{\rho, Y}
        +
        \left(
            \frac{\partial \ln \Gamma_1}{\partial \ln \rho}
        \right)_{P, Y}
    \right]
    \,
    K^{(\Gamma_1, \rrho)}
    +
    K^{(\rrho, \Gamma_1)}
\end{equation}
which gives
\begin{align}
K^{(\uu, Y)} &= 
\left({\partial \ln \Gamma_1 \over \partial \ln P} \right)_{\rho, Y}
	K^{(\Gamma_1, \rrho)}
	- \pp \left( {\psi \over \pp} \right)^{'} \\ 
	K^{(Y, \uu)} &= 
	\left({\partial \ln \Gamma_1 \over \partial Y} \right)_{\rho, P}
	K^{(\Gamma_1, \rrho)} \; _.
\end{align}

The kernel computation demonstrated that, for certain models,
$K^{(\uu, Y)}$ contains a very large smoothly varying component, apparently going through infinity and changing sign as the model evolves (see Figure~\ref{fig:kernel-evol}). 
Understanding this behavior requires analysis of the above transformations used to obtain the kernels. 
We start by expanding Equation~(\ref{eq:psieq}) to standard form as
\begin{equation} \label{eq:psieqex}
    \psi'' - {1 \over x} \left(2 + {\dd \ln \rrho \over \dd \ln x} \right) \psi' +
    	{4 \pi \rrho^2 \over \pp} \psi = 
    x^2 \rrho \left( { F \over x^2 \rrho } \right)'.
\end{equation}
At the singular point $x = 0$, we assume that $\psi \sim x^\alpha$ to leading order, and apply the Frobenius method to obtain the two solutions $\alpha = 3$ and $\alpha = 0$. 
To satisfy the boundary condition, $\alpha = 3$ must be selected, leading to a boundary condition for the numerical solution at the meshpoint closest to $x = 0$.%
\footnote{Since the two exponents differ by an integer, the solution corresponding to $\alpha = 0$ could include a logarithmic component; we have verified that this is not the case.}
However, for the analysis of the equation we consider both solutions $\psi_{\rm h,1}$ and $\psi_{\rm h,2}$ to the corresponding homogeneous equation, behaving respectively as $x^3$ and $x^0$ as $x \rightarrow 0$.
(We note that $\psi_{\rm h,1}$ does {\it not} in general satisfy the surface boundary condition $\psi_{\rm h,1}(1) = 0$.)
The general solution can then be written as
\begin{equation} \label{eq:psisol}
    \psi = \psi_{\rm p} + c_1 \psi_{\rm h,1} + c_2 \psi_{\rm h,2} 
\end{equation}
where $\psi_{\rm p}$ is a particular solution to the full equation. 
To determine a suitable $\psi_{\rm p}$, we first find the Wronskian of Equation~(\ref{eq:psieqex}) as
\begin{equation}
W \propto \exp\left[ \int^x {1 \over x} 
\left(2 + {\dd \ln \rrho \over \dd \ln x} \right)\, \dd x \right] 
	\propto x^2 \rrho.
\end{equation}
We choose the integration constant and the normalization of $\psi_{\rm h,1}$, $\psi_{\rm h,2}$ such that $W = x^2 \rrho$. Then $\psi_{\rm p}$ can be determined as
\begin{align}
    	\psi_{\rm p}(x) = \psi_{\rm h,1} &\int_x^1 \psi_{\rm h,2} 
    \left( { F \over s^2 \rrho } \right)'\, \dd s \notag
    	\\+ \psi_{\rm h,2} &\int_0^x \psi_{\rm h,1} 
    \left( { F \over s^2 \rrho } \right)'\, \dd s 
\end{align}
which is well-behaved at $x = 0$.
From the boundary condition at $x = 0$, $c_2 = 0$, and hence the boundary condition at $x = 1$ yields
\begin{equation}
c_1 = - 
	\frac{\psi_{\rm h,2}(1)}{\psi_{\rm h,1}(1)}
	\int_0^1 \psi_{\rm h,1} (x)
\left( { F \over x^2 \rrho } \right)' \, \dd x. 
\label{eq:psicoef}
\end{equation}
This is clearly singular if $\psi_{\rm h,1}(1) = 0$, resulting in the behavior that we have found for the computed kernels.

\begin{figure}
    \centering
    \includegraphics[width=\linewidth, trim={0 0 0 0.5cm}, clip]{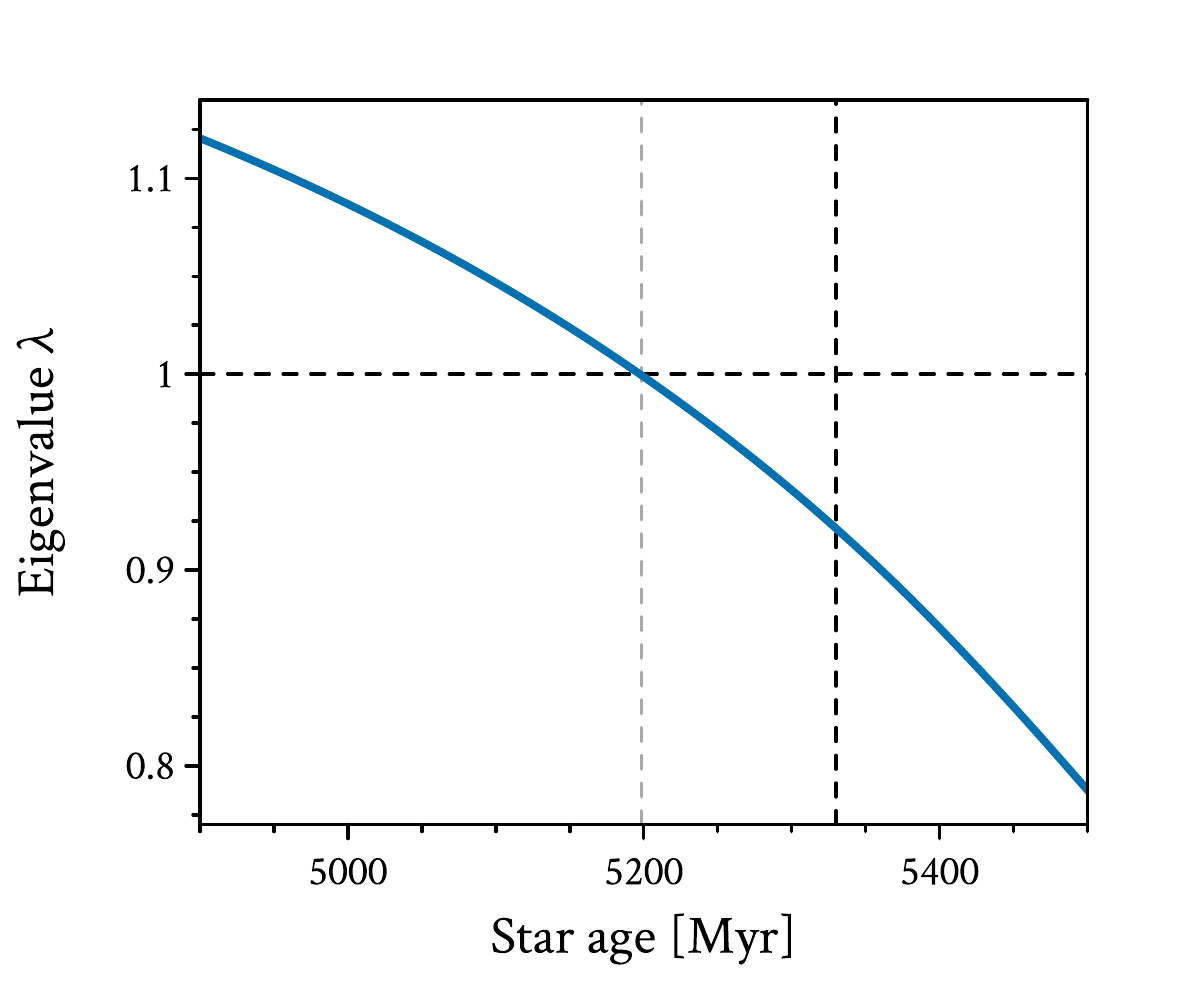}
    \caption{Eigenvalue $\lambda$ yielding a non-trivial solution to the homogeneous Equation~(\ref{eq:psieqlam}), against model age. 
    The horizontal line shows $\lambda = 1$. 
    Like Figure~\ref{fig:kernel-errs}, the gray vertical line indicates the age of the sign change and the black vertical line indicates the age of the first avoided crossing. 
    The model with $\lambda$ closest to 1, Model 1245, with age 5.199~Gyr, has $\lambda = 1.0001$. \label{fig:lambda}}
\end{figure}

\begin{figure}
    \centering
    \includegraphics[width=\linewidth, trim={0 0 0 0.5cm}, clip]{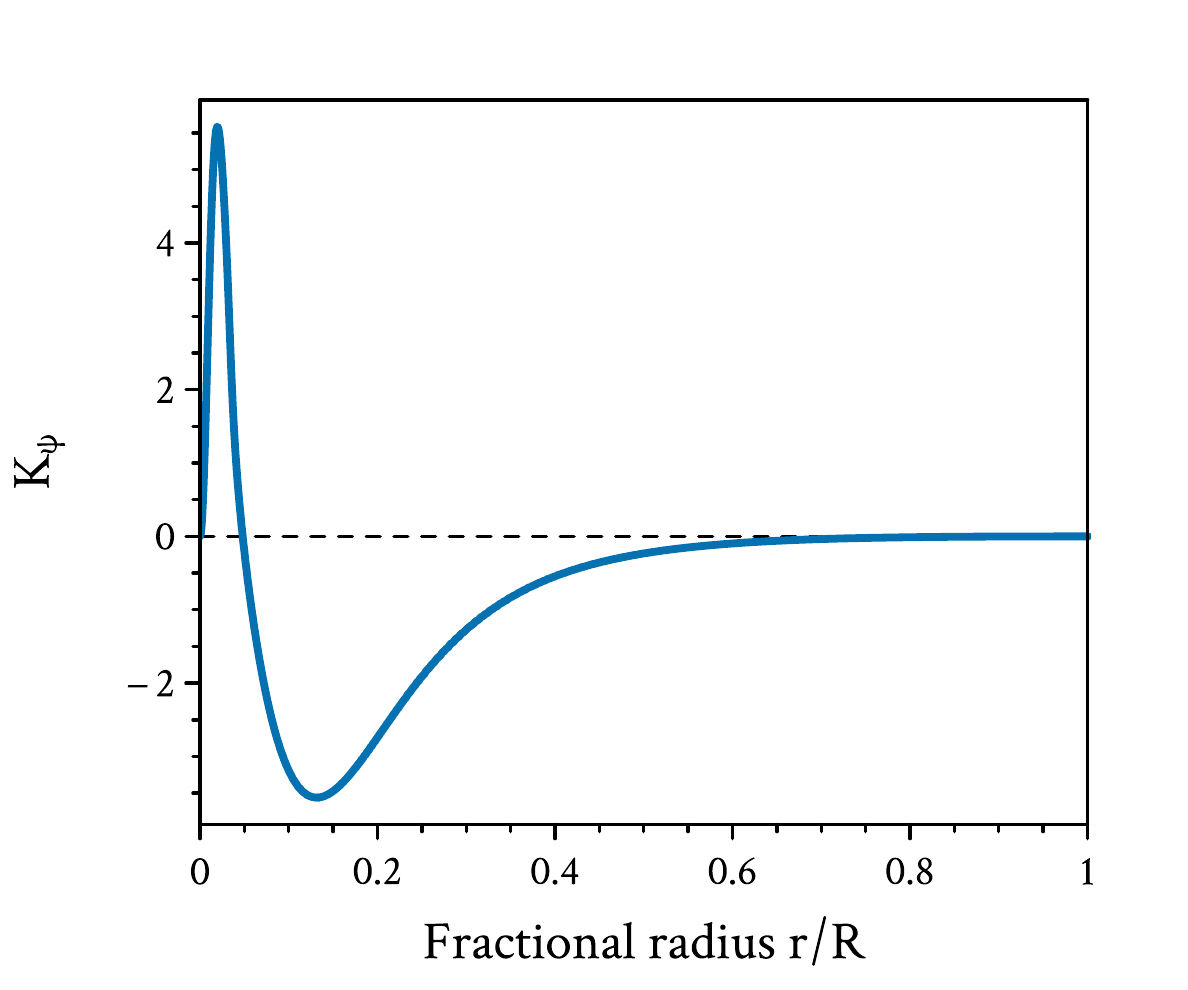}
    \caption{$K_\psi = \pp (\psi_\lambda /\pp)'$ determined from the eigenfunction corresponding to $\lambda = 1.0001$ in Model 1245 (see Figure~\ref{fig:lambda}). \label{fig:psilam}}
\end{figure}

To analyze this problem further, we consider the analogous homogeneous eigenvalue problem
\begin{equation} \label{eq:psieqlam}
    \left( 
        {\psi_\lambda' \over x^2 \rrho} 
    \right)' 
	+ 
	\lambda {4 \pi \rrho \over x^2 \hat P} \psi_\lambda 
	= 
	0 
\end{equation}
still with the boundary conditions $\psi_\lambda(0) = \psi_\lambda(1) = 0$.
Here $\lambda$ has been introduced as an eigenvalue which is determined such that the equation has a non-trivial solution, with the value of $\lambda$ obviously depending on the model. 
It is easy to show that as for the original equation there are two independent solutions, behaving as $x^3$ and $x^0$ for $x \rightarrow 0$.
It has been found numerically that Equation~(\ref{eq:psieqlam}) has a  series of eigenvalues.
The special case $\lambda = 1$ corresponds to the singular case, mentioned above, where $\psi_{\rm h,1}(1) = 0$.
Here we consider solutions to Equation~(\ref{eq:psieqlam}) with $\lambda$ close to 1, with corresponding eigenfunctions $\psi_\lambda$.
Figure~\ref{fig:lambda} shows the resulting behavior of $\lambda$ in the vicinity of the sign change in Figure~\ref{fig:kernel-evol}. 
Also, Figure~\ref{fig:psilam} shows $\pp (\psi_\lambda/\pp)'$, illustrating the corresponding slowly varying contribution to the kernels, for the model closest to $\lambda = 1$.
This model has a mass, radius and luminosity of $1.2~\text{M}_\odot$, $1.854~\text{R}_\odot$, $3.878~\text{L}_\odot$, and a helium core mass of $0.077~\text{M}_\odot$. 

\mbb{Clearly, models with ${\lambda=1}$ cannot serve as a reference model. Furthermore, comparing Figure~\ref{fig:lambda} with Figure~\ref{fig:kernel-errs} which shows the kernel errors when passing through the singularity, models with $\lambda - 1 \lessapprox 0.005$ may incur substantial errors, and thus a different reference model should be preferred. Models~A, B, C, and D have $\lambda = 1.115, 1.109, 1.108$, and $1.068$, respectively.} 

When $\lambda$ is close to 1 we can expand the homogeneous  solution $\psi_{\rm h,1}$ as
\begin{equation}
    \psi_{\rm h,1} = \psi_\lambda + \delta \psi .
\end{equation}
Here $\delta \psi$ satisfies, using the form in Equation~(\ref{eq:psieqex}),
\begin{equation}
    \delta \psi'' - 
    {1 \over x} \left(2 + {\dd \ln \rrho \over \dd \ln x} \right) \delta \psi' +
	\lambda {4 \pi \rrho^2 \over \hat P} \delta \psi = 
	(\lambda - 1) {4 \pi \rrho^2 \over \hat P} \psi_\lambda 
\label{eq:dpsieqex}
\end{equation}
where we neglected a term ${\cal O}(\delta \psi)$ on the right-hand side. Equation~(\ref{eq:dpsieqex}) has the same Wronskian as Equation~(\ref{eq:psieqex}) and hence the solution at $x = 1$ yields
\begin{equation}
	\psi_{h,1}(1) = \delta \psi(1) =
	(\lambda - 1) \psi_{\lambda,2}(1) 4 \pi \int_0^1
	{\rrho \over r^2 \hat P} \psi_\lambda^2 \, \dd x.
\end{equation}
Here $\psi_{\lambda,2}$ is the second solution to Equation~(\ref{eq:psieqlam}), behaving as $x^0$ as $x \rightarrow 0$. It follows from Eqs~(\ref{eq:psisol}) and (\ref{eq:psicoef}) that the slowly varying component of the kernel, corresponding to $\psi_{\rm h,1}$, scales as $(\lambda - 1)^{-1}$. This behavior is satisfied by the numerical kernels and corresponds to the sign change in Figure~\ref{fig:kernel-evol}.

Kernels multiplying $\delta \rrho/\rrho$ all involve a {\it complementary function} proportional to $x^2 \rrho$. This follows from the mass constraint, since 
\begin{equation}
	\int_0^1 x^2 \rrho {\delta \rrho \over \rrho}\, \dd x = 0.
\end{equation}
Consequently, such a kernel is only defined up to adding a multiple of $x^2 \rrho$. Inserting $x^2 \rrho$ for $F$ in Equation~(\ref{eq:psieq}) leads to the homogeneous equation; for a model with $\lambda = 1$ this has a non-trivial solution, which can then be identified with the complementary function for $\delta \uu/\uu$. Whether a complementary function for $\delta \uu/\uu$ exists for a general model is still unknown.
\section{Solving SOLA} \label{sec:minimize}

The task of minimizing Equation~(\ref{eq:sola}), suppressing also near-surface effects, leads to the system of linear equations \citep{1992A&A...262L..33P, basuchaplin2017}
\begin{align} 
    0 = \sum_j c_j\Bigg[
              &\int K_i^{(\hat u, Y)} K_j^{(\hat u, Y)} \,\text{d}x  \notag\\
    +\,  \beta &\int K_i^{(Y, \hat u)} K_j^{(Y, \hat u)} \,\text{d}x + \mu\, E_{i,j} \Bigg] \notag\\
    -\, &\int K_i^{(\hat u, Y)} \mathscr{T}\, \text{d}x + \lambda_1 \int K_i^{(\hat u, Y)} \text{d}x \notag\\
    +\, &\lambda_2 \left(\nu_i/\nu_\text{ac}\right)^2/I_i + \lambda_3 \left(\nu_i/\nu_\text{ac}\right)^{-2}/I_i
\end{align}
for $i=1,\ldots,M$, where $M$ is the number of observed modes. 
Here $\mathbf{E}$ denotes $\text{diag}(s_1^2, s_2^2, \ldots)$, $\boldsymbol\lambda$ are Lagrange multipliers, $\nu_{\text{ac}}$ is the acoustic cut-off frequency, and $\mathbf I$ are the mode inertias.
Additionally we have
\begin{equation}
    \sum_j c_j \int K^{(\hat u, Y)}_j \text{d}x = 1
\end{equation}
for unimodularity of the averaging kernel and 
\begin{align}
    \sum_j c_j \lambda_2 \left(\nu_i/\nu_\text{ac}\right)^2/I_i = 0 \\
    \sum_j c_j \lambda_3 \left(\nu_i/\nu_\text{ac}\right)^{-2}/I_i = 0
\end{align}
for the \citet{2014A&A...568A.123B} surface term. 
We minimize these equations to obtain $\mathbf c$ using the LDL$^T$ decomposition using the \textsc{Cholmod} package \citep{10.1145/1391989.1391995}. 
%

\bibliography{Bellinger}{}
\bibliographystyle{aasjournal}

\end{document}